\documentclass[12pt]{article}
\usepackage[margin=1.25in]{geometry}
\usepackage{amsthm}
\usepackage{amsmath}
\usepackage{amssymb}
\usepackage{amsfonts}
\usepackage{mathtools}
\usepackage{natbib}
\usepackage[colorlinks,citecolor=blue,urlcolor=blue,filecolor=blue,backref=page]{hyperref}
\usepackage{graphicx}
\usepackage[ruled]{algorithm2e}
\usepackage{float}
\usepackage{subfig}
\usepackage{comment}
\usepackage{bm}

\numberwithin{equation}{section}
\theoremstyle{plain}
\newtheorem{thm}{Theorem}[section]
\newtheorem{definition}{Definition}[section]

\newcommand{\ind}{\stackrel{\mathrm{ind}}{\sim}}

\DeclarePairedDelimiter\abs{\lvert}{\rvert}%

\begin{document}

\title{\large \bf Vector-weighted Mechanisms for Utility Maximization under Differential Privacy}

\author{Terrance D. Savitsky\footnote{U.S. Bureau of Labor Statistics, Office of Survey Methods Research, Suite 5930, 2 Massachusetts Ave NE Washington, DC 20212, Savitsky.Terrance@bls.gov}, Jingchen Hu\footnote{Vassar College, Box 27, 124 Raymond Ave, Poughkeepsie, NY 12604, jihu@vassar.edu} $\,$ and Matthew R. Williams\footnote{RTI International, 3040 East Cornwallis Road, Research Triangle Park, NC 27709, mrwilliams@rti.org}}

\maketitle

\begin{abstract}
We address practical implementation of a risk-weighted pseudo posterior synthesizer for microdata dissemination with a new re-weighting strategy that maximizes utility of released synthetic data under at any level of formal privacy guarantee. Our re-weighting strategy applies to any vector-weighted pseudo posterior mechanism under which a vector of observation-indexed weights are used to downweight likelihood contributions for high disclosure risk records.  We demonstrate our method on two different vector-weighted schemes that target high-risk records.  Our new method for constructing record-indexed downeighting  maximizes the data utility under any privacy budget for the vector-weighted synthesizers by adjusting the by-record weights, such that their individual Lipschitz bounds approach the bound for the entire database. Our method  achieves an $(\epsilon = 2 \Delta_{\bm{\alpha}})-$asymptotic differential privacy (aDP) guarantee, globally, over the space of databases. We illustrate our methods using simulated highly skewed count data and compare the results to a scalar-weighted synthesizer under the Exponential Mechanism (EM). We also apply our methods to a sample of the Survey of Doctorate Recipients and demonstrate the practicality of our methods.
\end{abstract}



\section{Introduction}
\label{intro}
Publishing survey and census data equipped with a privacy guarantee to limit the risk of respondent reidentificaiton is an important goal for government statistical agencies and private companies, worldwide. 
\subsection{Synthetic data generation to encode data privacy}
A commonly-used approach to encode privacy protection into data released to the public generates synthetic data from statistical models estimated on confidential, private data for proposed release by statistical agencies \citep{Rubin1993synthetic, Little1993synthetic}. This data synthesis approach replaces the confidential database with multiple synthetically generated record-level databases. The synthetic databases are released to the public who would use them to conduct any analyses of which they would conceive to be conducted on the real, confidential record-level data.  The synthetic data approach replaces multiple queries performed on a summary statistic with the publication of the synthetic databases encoded with privacy protection, making this approach independent of the specific queries performed by users or putative intruders.

\subsection{Differential privacy}
Our focus metric for measuring the relative privacy guarantee of our pseudo posterior synthesizing data mechanism introduced in the sequel is differential privacy \citep{Dwork:2006:CNS:2180286.2180305}.  We next provide a  definition for differential privacy \citep{McSherryTalwar2007}.

\begin{definition}[Differential Privacy]\label{def:DP}
Let $\mathbf{x}$ be a database in input space $\mathcal{X}^{n}$, where $\mathcal{X}^{n}$ denotes a space of databases of size (number of observations) $n$. Let $\mathcal{M}$ be a randomized mechanism  such that $\mathcal{M}(): \mathcal{X}^{n} \rightarrow O$. Then $\mathcal{M}$ is $\epsilon-$differentially private if
\[
\frac{Pr[\mathcal{M}(\mathbf{x}) \in O]}{Pr[\mathcal{M}(\mathbf{y}) \in O]} \le \exp(\epsilon),
\]
for all possible outputs $O = Range(\mathcal{M})$ under all possible pairs of datasets $\mathbf{x} \in \mathcal{X}^{n}$ where $\mathbf{y} \in \mathcal{X}^{n-1}$ differs from $\mathbf{x}$ by deleting one record or datum (under a leave-one-out (LOO) distance definition).
\end{definition}

Differential privacy is a property of the mechanism or data generating process and a mechanism that meets the definition above is guaranteed to be $\epsilon-$differentially private, or $\epsilon-$DP.  Differential privacy is called a ``formal" privacy guarantee because the $\epsilon$ level or guarantee is independent of the behavior of a putative intruder seeking to re-identify the data and the guarantee is not lessened by the existence of other data sources that may contain information about the same respondents included in $\mathcal{X}^{n}$.

Differential privacy assigns a disclosure risk for a statistic to be released to the public, $g(\mathbf{x})$ (e.g., total employment for a state-industry) of any $\mathbf{x} \in \mathcal{X}^{n-1}$ based on the global sensitivity, $\Delta = \mathop{\sup}_{\mathbf{x}\in\mathcal{X}^{n},\mathbf{y}\in\mathcal{X}^{n-1}: ~\delta(\mathbf{x},\mathbf{y})=1}|g(\mathbf{x}) - g(\mathbf{y})|$, over the space of databases, $\mathcal{X}^{n}$, where $\delta(\mathbf{x},\mathbf{y})$ denotes the number of records omitted from $\mathbf{x}$ in database, $\mathbf{y}$.  The distance metric, $\delta(\mathbf{x},\mathbf{y})$ denotes the LOO distance such that $\mathbf{x}$ differs from $\mathbf{y}$ by a single record, which is equivalent to using a Hamming-1 distance in the case of count based statistics of binary data records. If the value of the statistic, $g$, expresses a high magnitude change after the deletion of a data record in $\mathbf{y}$, then the mechanism will be required to induce a relatively higher level of distortion to $g$.  The more sensitive is a statistic to the change of a record, the higher its disclosure risk.

Our focus in this paper is where the mechanism, $\mathcal{M}$, is a model parameterized by $\theta$ from which replicate data are synthesized under an $\epsilon-$DP guarantee.  In particular, we leverage the smoothing property of $\theta$ to reduce the sensitivity and achieve a privacy guarantee in lieu of added noise to the confidential data distribution.  A common approach for generating parameter draws of $\theta$ under the statistical model for synthesizing data is the exponential mechanism (EM) of \citet{McSherryTalwar2007}, which inputs a non-private mechanism for $\theta$ and generates $\theta$ in such a way that induces an $\epsilon-$DP guarantee on the overall mechanism. The EM is conditioned on the availability of a global sensitivity over the space of databases, $\Delta_{u}$ for some utility function, $u(\mathbf{x}, \theta)$, defined on the space of databases and the space of parameters, globally. 

\subsection{Exponential mechanism for data synthesis}

\begin{definition}
(Exponential Mechanism)
The exponential mechanism releases values of $\theta$ from a distribution proportional to,
\begin{equation}
\exp \left(u(\mathbf{x}, \theta) \right)\xi\left(\theta\right),
\end{equation}
where $u(\mathbf{x}, \theta)$ is a utility function. Let \newline 
$\Delta_{u} = \mathop{\sup}_{\mathbf{x}\in \mathcal{X}^{n}} \,\, \mathop{\sup}_{\mathbf{x}, \mathbf{y}: \delta(\mathbf{x}, \mathbf{y}) = 1} \, \, \mathop{\sup}_{\theta \in \Theta} \,\, \abs{u(\mathbf{x}, \theta) - u(\mathbf{y}, \theta)}$ be the sensitivity, defined globally over $\mathbf{x} = (x_{1},\ldots,x_{n}) \in \mathcal{X}^n$, the $\sigma-$algebra of datasets, $\mathbf{x}$, governed by product measure, $P_{\theta_{0}}$ and the LOO distance metric, $\delta(\mathbf{x},\mathbf{y}) = 1$.  Then each draw of $\theta$ from the exponential mechanism is guaranteed to be $ \epsilon = 2\Delta_{u}-$DP.
\end{definition}
\noindent This result is based on the following definition of differential privacy under utility function, $u(\mathbf{x},\theta)$.
\begin{definition}
(Differential Privacy under the Exponential Mechanism)
A utility function, $u$, indexed by random parameters, $\theta$, gives $\epsilon-$DP if for all databases, $\mathbf{x} \in \mathcal{X}^{n}$ and associated databases, $\mathbf{y}:\delta(\mathbf{x}, \mathbf{y}) = 1$,
and all parameter values, $\theta \in\Theta$,
\begin{equation}
\mbox{Pr}\left(u(\mathbf{x},\theta) \in O\right)\leq \exp(\epsilon) \times \mbox{Pr}\left(u(\mathbf{y},\theta) \in O\right),
\end{equation}
where $O = \mbox{range}(u)$.
\end{definition}

In order to set an arbitrary $\epsilon \ne 2\Delta_{u}$, we must modify the utility function $u(\mathbf{x}, \theta)$. The statistical agency owning the confidential data will typically desire to determine $\epsilon$ as a matter of policy and not leave it to be $ \epsilon = 2\Delta_{u}$. The simplest and most common approach is to rescale it: $u^{*}(\mathbf{x}, \theta)=\frac{\epsilon}{2 \Delta_{u}}u(\mathbf{x}, \theta)$ \citep[See][among many others]{McSherryTalwar2007, Dwork:2006:CNS:2180286.2180305}.

The EM requires the availability of the sensitivity $\Delta_u$ for a chosen utility function $u(\mathbf{x}, \theta)$. \citet{WassermanZhou2010} and \citet{SnokeSlavkovic2018PSD} construct utility functions that are naturally bounded over all $\mathbf{x} \in \mathcal{X}^{n}$; however, they are not generally applicable to any population model and in the latter case are very difficult to implement in a computationally tractable manner since the EM distribution must be sampled by an inefficient random-walk Metropolis-Hastings scheme. 

For a Bayesian model utilizing the data log-likelihood as the utility function of the EM, \citet{SavitskyWilliamsHu2020ppm} demonstrate the EM mechanism becomes the model posterior distribution, which provides a straightforward mechanism from which to draw samples. \citet{Dimitrakakis:2017:DPB:3122009.3122020} define a model-based  sensitivity, \\ $ \mathop{\sup}_{\mathbf{x},\mathbf{y}\in \mathcal{X}^{n}:\delta(\mathbf{x}, \mathbf{y}) = 1}  \mathop{\sup}_{\theta \in \Theta} | f_{\theta}(\mathbf{x}) - f_{\theta}(\mathbf{y}) | \le \Delta$ that is constructed as a Lipschitz bound.  They demonstrate a connection between the Lipschitz bound, $\Delta$ and $\epsilon \leq 2\Delta$ for each draw of parameters, $\theta$, where $f_{\theta}(\mathbf{x})$ denotes the model log-likelihood indexed by $\theta$.  The guarantee applies to all databases $\mathbf{x}$, in the space of databases of size $n$, $\mathcal{X}^{n}$.

However, computing a finite $\Delta < \infty$ in practice, as acknowledged by \citet{Dimitrakakis:2017:DPB:3122009.3122020}, is difficult-to-impossible for an unbounded parameter space (e.g. a normal distribution) under simple models, which requires truncation of the parameter space to achieve a finite $\Delta$ and the truncation only works for some models to achieve a finite $\Delta$.  Moreover, parameter truncation becomes intractable for practical models that utilize a multidimensional parameter space.

\subsection{Pseudo posterior mechanism for data synthesis}

To guarantee the achievement of a  finite $\Delta < \infty$ for any synthesizing model over an unbounded parameter space, \citet{SavitskyWilliamsHu2020ppm} propose the \emph{pseudo} posterior mechanism that uses a log-pseudo likelihood with a vector of weights $\bm{\alpha} = (\alpha_1, \cdots, \alpha_n) \in [0,1]^{n}$ where each $\alpha_{i}$ exponentiates the likelihood contribution, $p(x_{i}\mid \theta)$, for each record $i \in (1,\ldots,n)$.  Each weight, $\alpha_{i} \in [0, 1]$ is set to be inversely proportional to a measure of disclosure risk for record, $i$, such that the model used to generate synthetic data will be less influenced by relatively high-risk records.  

The pseudo posterior mechanism of \citet{SavitskyWilliamsHu2020ppm} is formulated as
\begin{equation}
\label{pseudomech}
\xi^{\bm{\alpha}(\mathbf{x})}(\theta \mid \mathbf{x}) \propto \mathop{\prod}_{i=1}^{n}p(x_{i}\mid \theta)^{\alpha_{i}} \times \xi(\theta),
\end{equation}
where the $\alpha_{i} \in [0,1]$ serve to downweight the likelihood contributions with each $\alpha_{i} \propto 1/\mathop{\sup}_{\theta \in \Theta} \abs{f_{\theta}(x_{i})}$, where $f_{\theta}(x_{i}) = \log p(x_{i}\mid \theta)$ denotes the model log-likelihood, such that highly risky records are more strongly downweighted.  The differential downweighting of each record intends to better preserve utility by focusing the downweighting on high-risk records.  High-risk records tend to be those located in the tails of the distribution where the log-likelihood, $\abs{f_{\theta}(x_{i})}$, is highest, which allows the preservation of the high mass portions of the data distribution in the generated synthetic data.  The method sets $\alpha_{i} = 0$ for any record with a \emph{non-finite} log-likelihood, which \emph{ensures} a finite $\Delta_{\bm{\alpha}} = \mathop{\sup}_{\mathbf{x},\mathbf{y}\in \mathcal{X}^{n}:\delta(\mathbf{x}, \mathbf{y}) = 1}  \mathop{\sup}_{\theta \in \Theta} \lvert \alpha(\mathbf{x}) \times f_{\theta}(\mathbf{x}) - \alpha(\mathbf{y}) \times f_{\theta}(\mathbf{y}) \rvert < \infty$.  We see that $\Delta_{\bm{\alpha}} \leq \Delta$ since $\alpha_{i} \leq 1$.  
\begin{definition}\label{def:DP}
(Differential Privacy for the Pseudo Posterior Mechanism)
The  \\ $\bm{\alpha}-$weighted pseudo synthesizer, $\xi^{\bm{\alpha}(\mathbf{x})}(\theta \mid \mathbf{x})$, is a privacy mechanism defined in Equation~\ref{pseudomech}, which satisfies $\epsilon-$DP if the following inequality holds.
\begin{equation}
\mathop{\sup}_{\mathbf{x},\mathbf{y}\in \mathcal{X}^{n}:\delta(\mathbf{x}, \mathbf{y}) = 1} \mathop{\sup}_{B \in \beta_{\Theta}} \frac{\xi^{\bm{\alpha}(\mathbf{x})}(B \mid \mathbf{x})}{\xi^{\bm{\alpha}(\mathbf{y})}(B \mid \mathbf{y})} \leq e^{\epsilon}, 
\end{equation}
where $\xi^{\bm{\alpha}(\mathbf{x})}(B \mid \mathbf{x}) = \int_{\theta\in B}\xi^{\bm{\alpha}(\mathbf{x})}(\theta \mid \mathbf{x})d\theta$.
\end{definition}
 Definition~\ref{def:DP} limits the change in the pseudo posterior distribution over all sets, $B \in \beta_{\Theta}$ (i.e. $\beta_{\Theta}$ is the $\sigma-$algebra of measurable sets on $\Theta$), from the inclusion of a single record.   Although the pseudo posterior distribution mass assigned to $B$ depends on $\mathbf{x}$, the $\epsilon$ guarantee is defined as the supremum over all $\mathbf{x} \in \mathcal{X}^{n}$ and for all $ \mathbf{y}\in \mathcal{X}^{n}$ which differ by one record (i.e. $\delta(\mathbf{x},\mathbf{y})=1$).  
 
The $\alpha_{i}$ may not be released without leaking information because they are based / dependent on the confidential private data, $x_{i}$.  A draw of modeled parameters, however, may be released along with the synthetic data generated from those parameters (with no leakage of information since all that is released is synthetic data).
 
Let $\Delta_{\bm{\alpha},\mathbf{x}} = \mathop{\sup}_{\delta(\mathbf{x}, \mathbf{y}) = 1}  \mathop{\sup}_{\theta \in \Theta} \lvert \alpha(\mathbf{x}) \times f_{\theta}(\mathbf{x}) - \alpha(\mathbf{y}) \times f_{\theta}(\mathbf{y}) \rvert $ be the Lipschitz bound computed, locally, on database $\mathbf{x}$ (over all databases, $\mathbf{y}$, at a Hamming-1 distance from $\mathbf{x}$).  The pseudo posterior mechanism \emph{indirectly} sets the local DP guarantee, $\epsilon_{\mathbf{x}} = 2 \Delta_{\bm{\alpha},\mathbf{x}}$, through the computation of the likelihood weights, $\bm{\alpha}$.  

\citet{SavitskyWilliamsHu2020ppm} show that the local $\Delta_{\bm{\alpha},\mathbf{x}}$ contracts onto the global $\Delta_{\bm{\alpha}}$, asymptotically in sample size, which in turn drives the contraction of $\epsilon_{\mathbf{x}}$ onto $\epsilon$.  For a sample size $n$ sufficiently large, $\epsilon_{\mathbf{x}} = \epsilon$.  More formally, the authors demonstrate that the local Lipschitz satisfies a relaxed form of DP that they label ``asymptotic DP (aDP)".  

 We may imagine the generation of multiple collections of databases, $\{\bm{x}_{n,r}\}_{r=1}^{R}$, that produce the associated collection of Lipschitz bounds, $(\Delta_{\bm{\alpha},\bm{x}_{n,r}})_{r=1}^{R}$.  The aDP result guarantees that for $n$ sufficiently large, the local Lipschitz bounds for that collection of databases contract onto the global Lipschitz bound. In practice, \citet{SavitskyWilliamsHu2020ppm} show that in a Monte Carlo simulation study that for sample sizes of a few hundred that the $\Delta_{\bm{\alpha},\bm{x}_{n}} = \Delta_{\bm{\alpha}}$ up to any desired precision.  We conduct a Monte Carlo simulation study that generates multiple databases to illustrate the asymptotic convergence of a local privacy guarantee to a global privacy guarantee in the Supplementary Materials of this paper.

\subsection{Contribution of this paper}

\citet{SavitskyWilliamsHu2020ppm} provide a theoretical foundation for the pseudo posterior mechanism, and readers might be left with the question of how to set the $\bm{\alpha}$ weights, in practice.  While \citet{SavitskyWilliamsHu2020ppm} set each $\alpha_{i} \in [0,1]$ to be inversely proportional to the maximum (over $\theta$) absolute value of the log-likelihood for the record, there are possibly other ways to measure the disclosure risk for each record other than through the absolute value of the log-likelihood.  We will illustrate one alternative method for measuring risk used to set $\bm{\alpha}$ in the sequel. 

In this article, we focus on practical aspects of implementing alternative vector-weighted synthesizers, where each alternative synthesizer uses a \emph{different} approach for computing the weights.  The main contribution of this paper is to define a re-weighting strategy that inputs the vector of privacy weights, $\alpha_{i}$, formulated under any reasonable scheme that defines weights proportionally to the disclosure risks of the data records and subsequently adjusts those weights to achieve a maximally efficient weighting scheme under any level of an asymptotic DP privacy guarantee.  We use the word ``efficient" to denote the minimum distortion of the underlying distribution of the confidential data represented in the released synthetic data for a given privacy guarantee.

We propose a new re-weighting strategy that starts with computation of the maximum of the absolute value of log-pseudo likelihood values over the parameters sampled from the pseudo posterior synthesizer, $\Delta_{\bm{\alpha},x_{i}}$, for \emph{each} data record \emph{after} computing the weights, $\alpha_{i}$, and re-estimating the synthesizer under the $\bm{\alpha}-$weighted pseudo posterior model. The privacy guarantee is driven by the maximum over the data records, $x_{i},i \in (1,\ldots,n)$, of the $\Delta_{\bm{\alpha},x_{i}}$.  So any record, $i'$ with a $\displaystyle\Delta_{\bm{\alpha},x_{i'}} < \mathop{\max}_{i\in (1,\ldots,n)}\Delta_{\bm{\alpha},x_{i}} = \Delta_{\bm{\alpha},\mathbf{x}}$, that record is \emph{overly} downweighted since it does not determine the privacy protection for the overall database.  We scale up or increase these weight values, $\alpha_{i^{'}}$, for these overly downweighted data records in a linear re-weighting step that achieves the same formal privacy guarantee as under with the original weights, regardless of the weighting scheme used.  The re-weighting strategy improves the utility of the vector-weighted synthesizer while maintaining an equivalent privacy budget. The increased weights, in turn, reduce the distortion encoded for privacy into the released synthetic data, which improves its utility for the user.

The remainder of the article is organized as follows. Section \ref{dp} describes the key steps and the implementation algorithms of two vector-weighted synthesizers under different schemes to define privacy weights, $\bm{\alpha}$. We introduce the new re-weighting strategy in Section \ref{reweight} with an algorithm and a simulation study of highly skewed count data applied to both synthesizers. A Monte Carlo simulation study to demonstrate contraction of local Lipschitz values onto a global Lipschitz is included in the Supplementary Materials. We apply our methods to the highly skewed salary variable from a sample of the Survey of Doctorate Recipients in Section and \ref{app}. Section \ref{conclusion} ends this article with a few concluding remarks.

\section{Vector-weighted Synthesizer Algorithms}
\label{dp}

The second synthesizer, labeled as count-weighted (CW), sets each by-record weight, $\alpha_i \in [0,1]$, such that $\alpha_i \propto 1 / IR_i$, where $IR_i$ denotes the disclosure risk probability ($\in [0,1]$) of record $i$. The $IR_i$ is a measure of a record's isolation from other records and is constructed by counting the number of records whose values are outside a radius around the true value for the target record divided by the total number of records \citep{HuSavitskyWilliams2020rebds}. A record whose true value is not well-covered by the values of other records is relatively more isolated and, therefore, at higher disclosure risk.  The radius is a measure of closeness that is tuned by the owner of the confidential data.  

In both synthesizers, the record-indexed vector weights $\bm{\alpha} = (\alpha_1 \in [0,1], \cdots, \alpha_n \in [0,1])$ are used to exponentiate the likelihood contributions where the weights are designed to target high-risk records by downweighting their likelihood contributions.  These two measures of risk, LW and CW, are related in that the notion of record isolation underlies both.   The value of the response variable for an isolated target record is near to or within a close radius to the values of many other records.  As earlier mentioned, such isolated records generally appear in the tails of the distribution where there is little distribution mass.  As a result, downweighting the likelihood contribution of isolated records tends to preserve the high mass regions of the confidential data distribution in the resulting synthetic data generated for release.

We specify the method for formulation of vector weights $\bm{\alpha} = (\alpha_{1},\ldots,\alpha_{n})$ for the LW and CW synthesizers in Section \ref{dp:PPM} and Section \ref{dp:count}, with Algorithm \ref{al:all3-PPM} and Algorithm \ref{al:all3-count}, respectively.

Each algorithm starts by computing the weights, $\bm{\alpha} = (\alpha_{1},\ldots,\alpha_{n})$, which are then used to construct the pseudo likelihood and estimate the pseudo posterior. Next, we draw parameters from the estimated pseudo posterior distribution and compute the overall Lipschitz bound, $\Delta_{\bm{\alpha}, \mathbf{x}}$ for database, $\mathbf{x}$.  The resulting DP guarantee is $(\epsilon_{\mathbf{x}} = 2\Delta_{\bm{\alpha}, \mathbf{x}})$ and is ``local" to the database, $\mathbf{x}$, and the $\epsilon_{\mathbf{x}}$ is \emph{indirectly} controlled through the weights, $\bm{\alpha}$.  Synthetic data are then generated using the drawn $(\theta_{s})_{s=1,\ldots S}$ from the $\bm{\alpha}-$weighted pseudo posterior distribution from step 5 in each algorithm of the corresponding data generating model. 

As earlier discussed, the local Lipschitz bound, $\Delta_{\bm{\alpha}, \mathbf{x}}$, contracts onto the ``global" Lipschitz bound, $\Delta_{\bm{\alpha}}$, over all databases, $\mathbf{x}\in \mathcal{X}^{n}$ of size $n$, as $n$ increases such that $\epsilon_{\mathbf{x}}$ contracts onto $\epsilon$ at a relatively modest sample size.

\subsection{Generating synthetic data under the LW  synthesizer}
\label{dp:PPM}

We specify the algorithm for generating synthetic data under the LW $\bm{\alpha}-$weighted pseudo posterior distribution that produces synthetic data for database, $\mathbf{x}$.

To implement the LW vector-weighted synthesizer, we first fit an unweighted synthesizer and obtain the absolute value of the log-pseudo likelihood for each data base record $i$ and each Markov chain Monte Carlo (MCMC) draw $s$ of $\theta$ from the unweighted posterior distribution. A Lipschitz bound for each record is computed by taking the maximum of the log-likelihoods over the $S$ posterior draws of $\theta$. We formulate by-record weights, $\bm{\alpha} = (\alpha_1, \cdots, \alpha_n)$, to be inversely proportional to those by-record Lipschitz bounds. See step 1 to step 4 in Algorithm \ref{al:all3-PPM}.

\begin{algorithm}[t]
\SetAlgoLined
1. Let $\lvert f_{\theta_{s},i} \rvert$ denote the absolute value of the log-likelihood computed from the unweighted synthesizer for database record, $i \in (1,\ldots,n)$ and MCMC draw, $s \in (1,\ldots,S)$ of $\theta$ from its unweighted posterior distribution\;
2. Compute the $S \times n$ matrix of by-record absolute value of log-likelihoods, $L = \left\{ \lvert f_{\theta_{s},i} \rvert \right\}_{i=1,\ldots,n,~s=1,\ldots,S}$\;
3. Compute the maximum over each $S\times 1$ column of $L$ to produce the $n\times 1$ (database record-indexed) vector, $\mathbf{f} = \left(f_{1},\ldots,f_{n}\right)$.  We use a linear transformation of each $f_{i}$ to $\tilde{f}_{i} \in [0,1]$ where values of $\tilde{f}_{i}$ closer to $1$ indicates relatively higher identification disclosure risk:
$\tilde{f}_{i} = \frac{f_i - \min_j f_j}{\max_j f_j - \min_j f_j}$\;
4. Formulate by-record weights, $\bm{\alpha} = (\alpha_1, \cdots, \alpha_n)$, $\alpha_i =  c \times (1-\tilde{f}_{i}) + g$, where $c$ and $g$ denote a scaling and a shift parameters, respectively, of the $\alpha_i$ used to tune the risk-utility trade-off for setting $\epsilon_{\mathbf{x}} = 2\Delta_{\bm{\alpha},\mathbf{x}}$\;
5. Use $\bm{\alpha} = \left(\alpha_{1},\ldots,\alpha_{n}\right)$ to construct the $\bm{\alpha}-$weighted pseudo posterior distribution, $\xi^{\bm{\alpha}(\mathbf{x})}(\theta \mid \mathbf{x}) \propto \mathop{\prod}_{i=1}^{n}p(x_{i}\mid \theta)^{\alpha_{i}} \times \xi(\theta)$.
 Draw $(\theta_{s})_{s=1,\ldots S}$ from the $\bm{\alpha}-$weighted pseudo posterior distribution, where $S$ denotes the number of draws of $\theta$ from the $\bm{\alpha}-$weighted pseudo posterior distribution\;
6. Compute the $S\times~n$ matrix of log-pseudo likelihood values, $L^{\bm{\alpha}} = \left\{ \lvert f^{\alpha_{i}}_{\theta_{s},i} \rvert \right\}_{i=1,\ldots,n,~s=1,\ldots,S}$ where $f^{\alpha_{i}}_{\theta_{s},i} = \log p(x_{i}\mid \theta_{s})^{\alpha_{i}}$\;
7. Compute $\Delta_{\bm{\alpha},\mathbf{x}} = \mathop{\max}_{s,i} \lvert f^{\alpha_{i}}_{\theta_{s},i} \rvert$, that defines the local DP guarantee, $\epsilon_{\mathbf{x}} = 2\Delta_{\bm{\alpha},\mathbf{x}}$, for database $\mathbf{x}$.
 \caption{Steps to implement the LW vector-weighted synthesizer}
 \label{al:all3-PPM}
\end{algorithm}

Algorithm \ref{al:all3-PPM} is implemented on the observed database, $\mathbf{x}\in\mathcal{X}^n$, under which we compute the local (specific-to-database $\mathbf{x}$) Lipschitz bound, $\Delta_{\bm{\alpha},\mathbf{x}}$, to achieve a local privacy privacy guarantee, $\epsilon_{\mathbf{x}} = 2\Delta_{\bm{\alpha},\mathbf{x}}$, which is equivalent to an $\epsilon =\epsilon_{\mathbf{x}}-$aDP guarantee where $\epsilon = \epsilon_{\mathbf{x}}$ for $n$ sufficiently large. 

We emphasize that LW \emph{indirectly} achieves the aDP guarantee, $(\epsilon = 2\Delta_{\mathbf{\alpha}},\delta)$, through the computation of the likelihood weights, $\bm{\alpha}$. Sample R scripts implementing Algorithm \ref{al:all3-PPM} are available in the Supplementary Materials. 

The LW algorithm constructs weights intended to directly minimize the overall Lipschitz bound for the synthetic data by downweighting the likelihood contribution for each record inversely proportional to how large is the absolute value of its log-likelihood.  We recall that the Lipschitz is the maximum over the parameter space and records of this absolute value of the log-likelihood quantity, so our LW weighting scheme will be efficient at targeting those high risk records that most effect the privacy guarantee to produce a relatively moderate distortion of the confidential, confidential data distribution expressed in the publicly-released synthetic data.

\subsection{Generating synthetic data under the CW synthesizer}
\label{dp:count}
Next, we present the algorithm of \citet{HuSavitskyWilliams2020rebds}
for generating synthetic data under the CW $\bm{\alpha}-$weighted synthesizer.
The weights $\bm{\alpha}$ are estimated as probabilities of identification disclosure, and each $\alpha_{i} \in [0,1]$, based on the assumption that a putative intruder guesses randomly from a collection of records whose values are near to or within some set radius of the record being identified.

To compute a weight for each record, $i \in (1,\ldots,n)$, we first calculate its estimated probability of identification disclosure. We assume that an intruder knows the data value of the record she seeks and that she will randomly choose among records that are close to that value. More formally, we cast a ball, $B(y_i, r)$, around the true value of $y_i$ for record $i$ with a radius $r$. The radius, $r$, is a policy hyperparameter set by the agency who owns the confidential data.  We count the number of records whose values fall \emph{outside} of the radius around the target, and take the ratio of this count over the total number of records, a proportion that we label the risk probability of identification. A target record where the values for most other records lie outside the radius are viewed as isolated because the target record value is sparsely covered by the values of other records, and therefore at a higher risk of identification disclosure.  We then formulate by-record weights, $\bm{\alpha} = (\alpha_1, \cdots, \alpha_n)$, that are inversely proportional to the by-record risk probabilities. See step 1 to step 4 in Algorithm \ref{al:all3-count}.

Even though the weights under CW are computed based on assumptions about the intruder behavior, we are yet able to compute its $(\epsilon_{\mathbf{x}} = 2\Delta_{\bm{\alpha},\mathbf{x}})$ and invoke the aDP guarantee of \citet{SavitskyWilliamsHu2020ppm} since any weighting scheme with $\bm{\alpha} \in [0,1]$ produces this formal privacy guarantee under mild conditions that regulate $\bm{\alpha}$. 

\begin{algorithm}[t]
\SetAlgoLined
1. Let $M_i$ denote the set of records in the original data, and $\lvert M_i \rvert$ denote the number of records in the set\;
2. Cast a ball, $B(y_i, r)$ with a radius $r$ around the true value of record $i$, and count the number of records falling outside the radius $\sum_{j \in M_i} \mathbb{I}\left(y_j \notin B(y_i, r)\right)$\;
3. Compute the record-level risk probability, $IR_i$ as $IR_i = \sum_{j \in M_i} \mathbb{I}\left(y_j \notin B(y_i, r)\right) / \lvert M_i \rvert$, such that $IR_i \in [0, 1]$\;
4. Formulate by-record weights, $\bm{\alpha} = (\alpha_1, \cdots, \alpha_n)$, $\alpha_i =  c \times (1-IR_i) + g$, where $c$ and $g$ denote a scaling and a shift parameters, respectively, of the $\alpha_i$ used to tune the risk-utility trade-off\;
5. Use $\bm{\alpha} = \left(\alpha_{1},\ldots,\alpha_{n}\right)$ to construct the pseudo likelihood from which the pseudo posterior is estimated.  Draw $(\theta_{s})_{s=1,\ldots S}$ from the $\bm{\alpha}-$weighted pseudo posterior distribution\;
6. Compute the $S\times~n$ matrix of log-pseudo likelihood values, $L^{\bm{\alpha}} = \left\{ \lvert f^{\alpha_{i}}_{\theta_{s},i} \rvert \right\}_{i=1,\ldots,n,~s=1,\ldots,S}$\;
7. Compute $\Delta_{\bm{\alpha},\mathbf{x}} = \mathop{\max}_{s,i} \lvert f^{\alpha_{i}}_{\theta_{s},i} \rvert$, that defines the local DP guarantee for database $\mathbf{x}$.
 \caption{Steps to implement the CW vector-weighted synthesizer.}
 \label{al:all3-count}
\end{algorithm}
\vspace{1mm}


\section{Re-weighting to Maximize Utility for \emph{Any} Vector-weighted Synthesizer}
\label{reweight}

\subsection{Motivation and the proposed method}
\label{reweight:method}

To motivate our re-weighting strategy we simulate data from a mixture of two negative binomial distributions, $\textrm{NB}(\mu = 100, \phi = 5)$ and $\textrm{NB}(\mu = 100, \phi = 20)$, where $\phi$ denotes an over-dispersion parameter under which the variance is allowed to be larger than the mean  (with mixture weights of $\pi = 0.2$ and $(1-\pi) = 0.8$), which produces data with a highly skewed distribution. All model estimations are performed in Stan \citep{Rstan}. The Stan script for a weighted negative binomial synthesizer is available in the Supplementary Materials.

We label the vector-weighted synthesizer in Section \ref{dp:PPM} as LW, and that in Section \ref{dp:count} as CW. For comparison, we include a scalar-weighted synthesizer with a scalar weight for every record set as $\alpha_i = \Delta_{\mathbf{\alpha},\mathbf{x}} / \Delta_{\mathbf{x}}$ for all $i \in (1,\ldots,n)$, where $\Delta_{\mathbf{x}}$ is the local Lipschitz bound for the \emph{unweighted} synthesizer. This scalar-weighted pseudo posterior synthesizer has been shown equivalent to the EM under a log-likelihood utility and produces an $\epsilon_{\mathbf{x}} = 2\Delta_{\mathbf{\alpha},\mathbf{x}}-$aDP guarantee \citep{SavitskyWilliamsHu2020ppm}. This allows us to set the $\epsilon_{\mathbf{x}}$ for the scalar-weighted synthesizer (SW) to that for the LW and the CW vector-weighted synthesizers to compare their relative utility performances at equivalent level of privacy protection.  Finally, we include the unweighted synthesizer, labeled as ``Unweighted", which is negative binomial synthesizer for the mixture of negative binomial-simulated data.

\begin{figure}[t]
  \centering
  \includegraphics[width=1\textwidth]{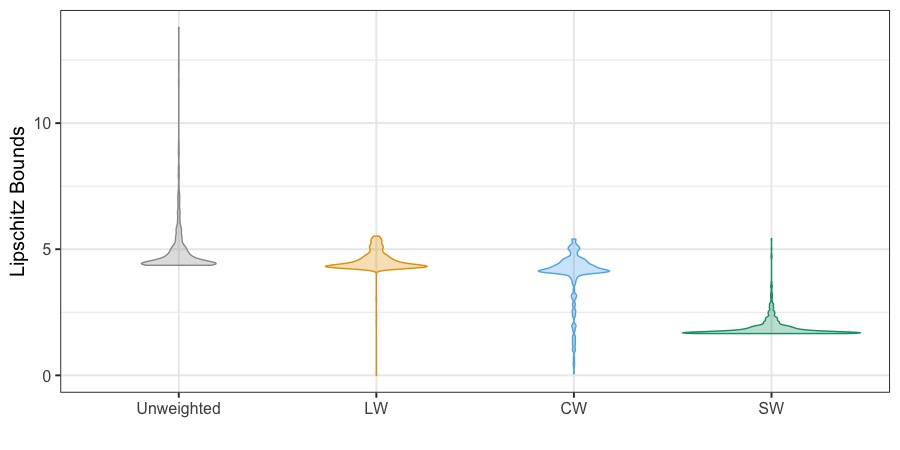}
  \caption{Distributions of record-level Lipschitz bounds of the unweighted synthesizer, the LW and CW vector-weighted synthesizers and the EM synthesizer. All weighted synthesizers create maximum Lipschitz bounds that are significant lower than the Unweighted. Among them, the LW shows the best control of the distribution of Lipschitz bounds with the majority of the by-record Lipschitz bounds being closer to the maximum.}
  \label{fig:NB-risks}
 \end{figure}

Figure~\ref{fig:NB-risks} presents the by-record Lipschitz bounds the two vector-weighted synthesizers, LW and CW. We know that the $(\epsilon_{\mathbf{x}} = 2\Delta_{\bm{\alpha},\mathbf{x}})-$aDP privacy guarantee is controlled by the maximum Lipschitz bound $\Delta_{\bm{\alpha},\mathbf{x}}$. As long as the maximum of the by-record Lipschitz bounds remains unchanged, we may increase the by-record Lipschitz bounds for other records whose bounds lie below the maximum value to be closer to the overall maximum value without any loss of privacy since the aDP guarantee is based on the maximum Lipschitz bound among the records.


\begin{figure}[t]
  \centering
    \includegraphics[width=1\textwidth]{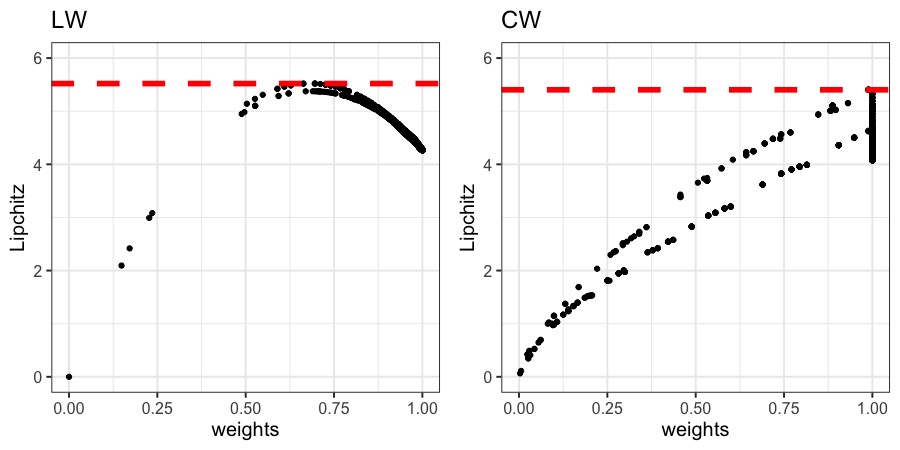}
    \caption{Lipschitz Bounds vs Weights, LW (left) and CW (right). The LW creates the majority of records expressing high Lipschitz bounds due to high weight values and therefore more efficient. The CW produces many records with low Lipscthiz bounds due to their over-downweighting.}
    \label{fig:NB-L-alpha}
\end{figure}


Figure \ref{fig:NB-L-alpha} shows scatter plots of by-record Lipschitz bounds $\Delta_{\bm{\alpha},x_{i}}$ (on the y-axis) against by-record weights, $\alpha_i$, (on the x-axis) for LW and CW. In each case, the red dashed line indicates the maximum Lipschitz bound $\Delta_{\bm{\alpha},\mathbf{x}}$. Only a small number of records express low Lipschitz bounds for the LW synthesizer and the majority  of records express relatively high Lipschitz bounds due to high weight values. By contrast, CW produces many records with low Lipschitz bounds due to their over-downweighting. The relative concentration of higher by-record weights and associated Lipschitz bounds for LW than for CW serve as a further justification of why LW is a more ``efficient" vector-weighted scheme than CW.  In particular, LW has done a relatively good job of targeting high-risk records and down-weighting them, while CW is overly downweighting lower-risk records and is less targeted than LW (to achieve the same privacy guarantee).

Moreover, Figure \ref{fig:NB-L-alpha} reveals that both vector-weighted synthesizers could improve their weighting efficiency to achieve a given maximum Lipschitz bound $\Delta_{\bm{\alpha},\mathbf{x}}$. We can increase the weights so that the by-record Lipschitz bounds $\Delta_{\bm{\alpha}, x_i}$'s become closer to the red dashed line. In this way, we can improve the utility performance of LW and CW while maintaining an equivalent overall $\Delta_{\bm{\alpha},\mathbf{x}}$. The utility will improve because less downweighting of records produces less distortion of the confidential data distribution in the generated synthetic data.  In the limit, the best efficiency that may be achieved by a vector-weighted scheme is one where the plot of by-record weights on the x-axis and the by-record Lipschitz bounds on the y-axis is horizontal; that is, there is no relationship between the weights and the Lipschitz bounds.

Our re-weighting strategy constructs re-weighted weights $\bm{\alpha}^{w} = (\alpha_1^{w}, \cdots, \alpha_n^{w})$ by:
\begin{equation}
\alpha_i^{w} = k \times \alpha_i \times \frac{\Delta_{\bm{\alpha},\mathbf{x}}}{\Delta_{\bm{\alpha},x_i}},
\end{equation}
where $\Delta_{\bm{\alpha},\mathbf{x}}$ is the maximum Lipschitz bound, and $\Delta_{\bm{\alpha},x_i}$ is the Lipschitz bound for record $i$. $\alpha_i$ is the weight used in the pseudo posterior synthesizers before the re-weighting step, and a constant, $k < 1$, is used to ensure that the final maximum Lipschitz bound remains equivalent before and after this re-weighting step. Both $\Delta_{\bm{\alpha},\mathbf{x}}$ and $\Delta_{\bm{\alpha},x_i}$ are computed from the $\bm{\alpha}-$weighted pseudo posterior before re-weighting.  The implementation for the new re-weighting step is outlined in Algorithm \ref{al:reweight}, and should be inserted between step 4 and step 5 in Algorithm \ref{al:all3-PPM} and Algorithm \ref{al:all3-count} for LW and CW, respectively.

\begin{algorithm}
\SetAlgoLined
1. Use the calculated $\bm{\alpha} = (\alpha_1, \cdots, \alpha_n)$ from the unweighted synthesizer.  Use the overall Lipschitz bound, $\Delta_{\bm{\alpha},\mathbf{x}}$, and the by record Lipschitz bounds, $\{\Delta_{\bm{\alpha},x_i}, i = 1, \cdots, n\}$, computed from the $\bm{\alpha}-$weighted pseudo posterior synthesizer.  Construct a constant $k < 1$ to compute $\bm{\alpha}^{w}$, where each $\alpha_i^{w} = k \times \alpha_i \times \frac{\Delta_{\bm{\alpha},\mathbf{x}}}{\Delta_{\bm{\alpha},\mathbf{x}_i}} \in [0,1]$.\;
2. Run step 5 to 7 in Algorithm \ref{al:all3-PPM} / Algorithm \ref{al:all3-count}, again, to re-estimate the synthesizers under an $\bm{\alpha}^{w}-$weighted pseudo posterior to obtain $\Delta_{\bm{\alpha}^{w},\mathbf{x}}$ to make sure that $\Delta_{\bm{\alpha},\mathbf{x}} \approx \Delta_{\bm{\alpha}^{w},\mathbf{x}}$. If not, try another $k < 1$ and repeat.
 \caption{Re-weighting step to obtain $\bm{\alpha}^{w}$}
  \label{al:reweight}
\end{algorithm}

\subsection{Application to simulated highly skewed data}
\label{reweight:app}


We demonstrate our re-weighting strategy under the highly skewed negative binomial mixture data. Using $k = 0.95$ produces an equivalent overall Lipschitz bound. The re-weighted synthetic data results are labeled ``LW\_final" and ``CW\_final", for LW and CW respectively.

\begin{figure}[t]
    \centering
    \subfloat[LW]{    \includegraphics[width=0.48\textwidth]{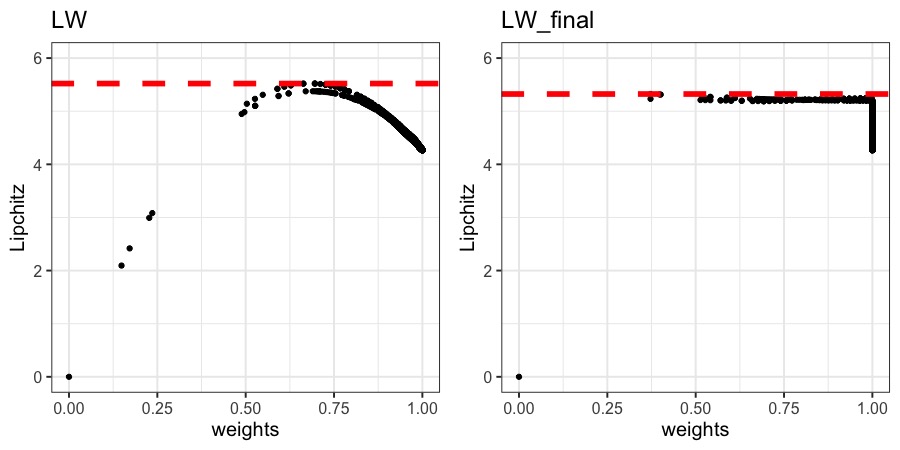}
    \label{fig:alpha-L-LW}}
    \subfloat[CW]{    \includegraphics[width=0.48\textwidth]{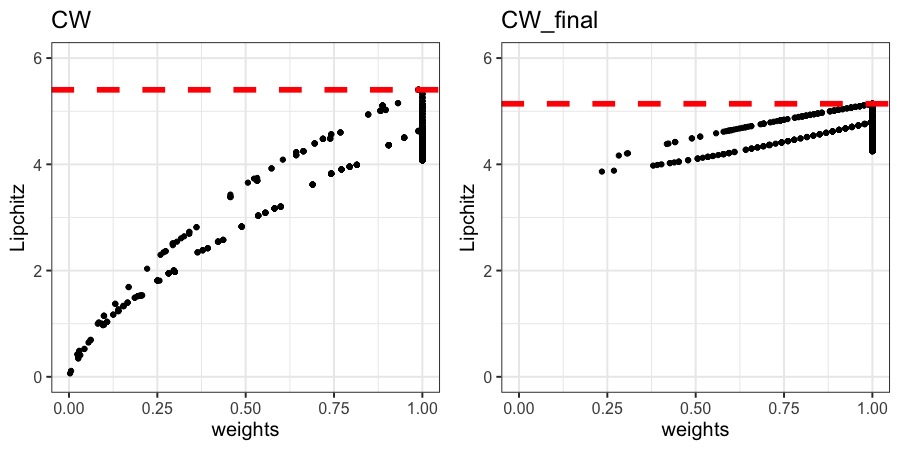}
    \label{fig:alpha-L-CW}}
    \caption{Lipschitz Bounds vs Weights, before and after re-weighting for LW (left) and CW (right). The re-weighting strategy produces a nearly horizontal Lipschitz-weight relationship. The impact is more significant for the LW.}
    \label{fig:alpha-L}
\end{figure}

Figure \ref{fig:alpha-L-LW} and Figure \ref{fig:alpha-L-CW} show the before vs after re-weighting scatter plots of Lipschitz bounds and weights. As is evident in Figure \ref{fig:alpha-L-LW}, the curve showing the Lipschitz-weight relationship becomes nearly horizontal at the maximum Lipschitz bound $\Delta_{\bm{\alpha},\mathbf{x}}$ as we move from LW to LW\_final, which indicates maximum efficiency. The curve in Figure \ref{fig:alpha-L-CW} becomes much less vertical from CW to CW\_final, indicating much improved efficiency.
%

\begin{figure}[t]
    \centering
    \subfloat[Lipschitz Bounds]{    \includegraphics[width=0.48\textwidth]{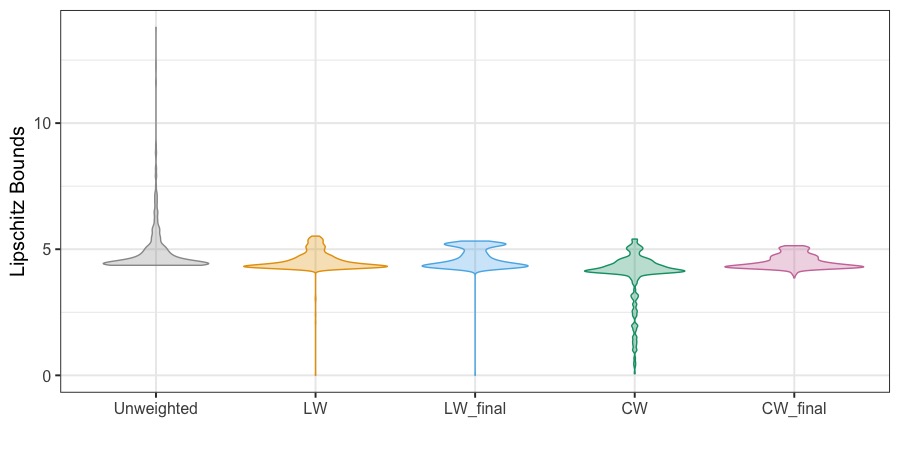}
    \label{fig:NB-risks-2}}
    \subfloat[Record-level Weights]{    \includegraphics[width=0.48\textwidth]{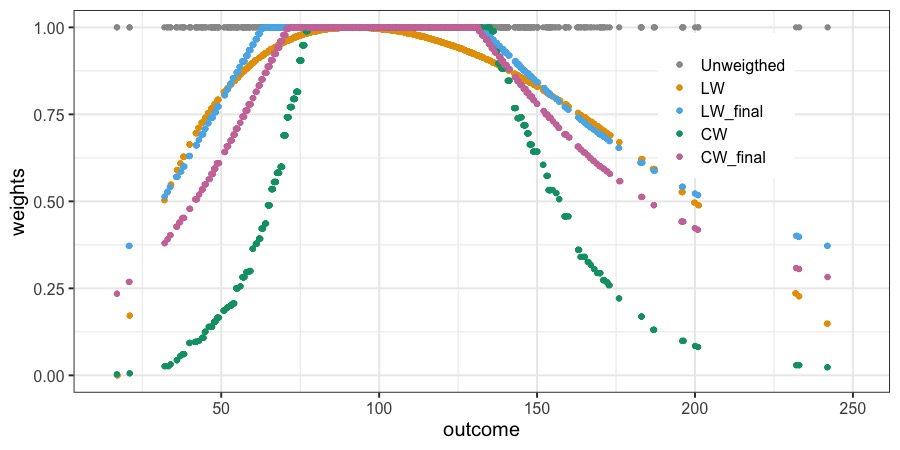}
    \label{fig:NB-weights-dots-2}}\\
    \subfloat[LW Weights]{    \includegraphics[width=0.48\textwidth]{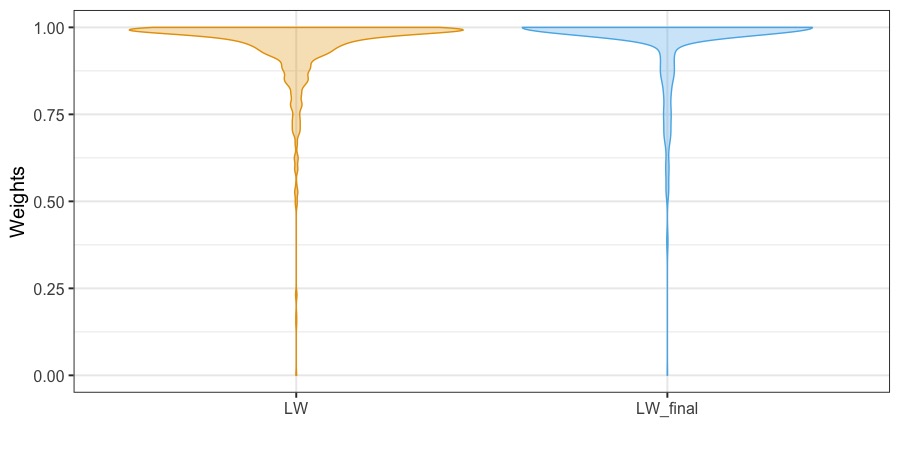}
    \label{fig:NB-weights-LW}}
    \subfloat[CW Weights]{    \includegraphics[width=0.48\textwidth]{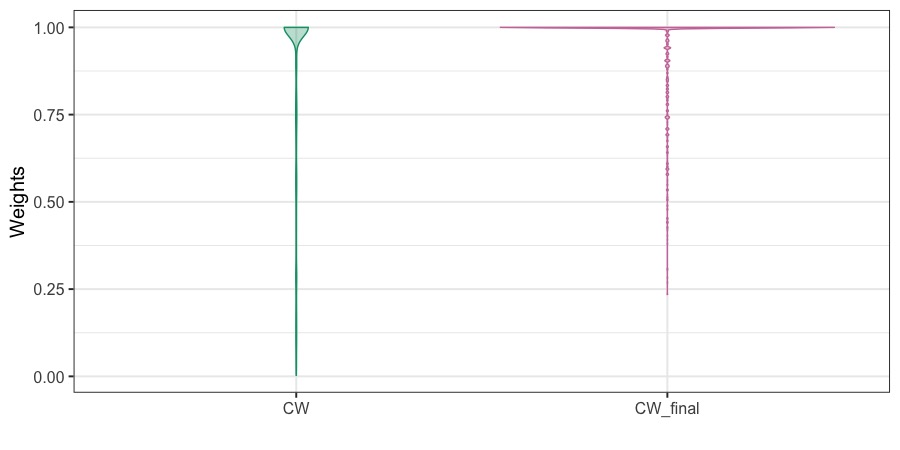}
    \label{fig:NB-weights-CW}}
    \caption{Before vs after re-weighting for the negative binomial mixture: Lipshitz Bounds (panel a) and Weights (panels b to d). After re-weighting, the by-record Lipschitz bounds are closer to the maximum, whereas the record-level weights are closer to one, indicating the effectiveness of the re-weighting strategy to increase the weighting efficiency of both LW and CW.}
    \label{fig:NB-reweight-weights}
\end{figure}

Figure \ref{fig:NB-risks-2} confirms that our re-weighting strategy increases the weighting efficiency of LW and CW in that the by-record Lipschitz bounds are increased while maintaining an equivalent maximum Lipschitz bound.  At the record level, Figure \ref{fig:NB-weights-dots-2} illustrates that every record has received a higher weight from LW to LW\_final, and from CW to CW\_final. We receive further confirmation of the improved efficiency of implementing the re-weighting step from the weight plots in Figure \ref{fig:NB-weights-LW} and Figure \ref{fig:NB-weights-CW} that show weights increase after re-weighting.

\begin{figure}[t]
    \centering
    \subfloat[15th and 90th Quantiles]{    \includegraphics[width=0.48\textwidth]{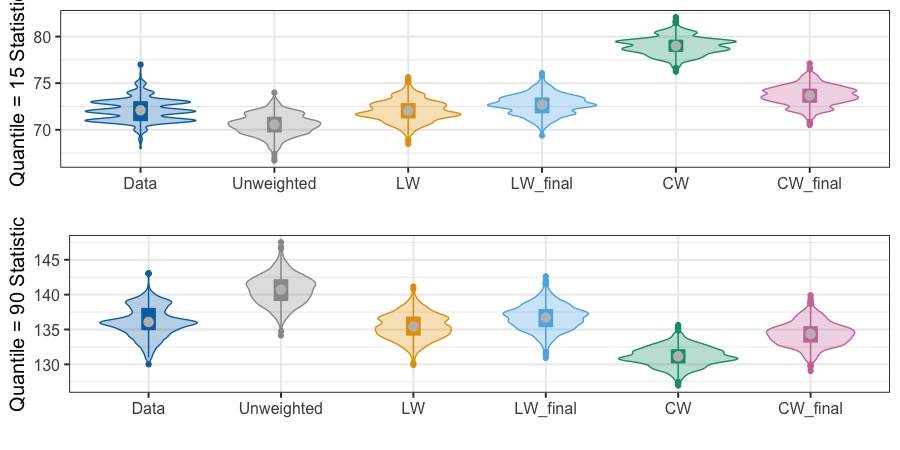}
      \label{fig:NB-utility4-reweight}}
    \subfloat[Mean and Median]{    \includegraphics[width=0.48\textwidth ]{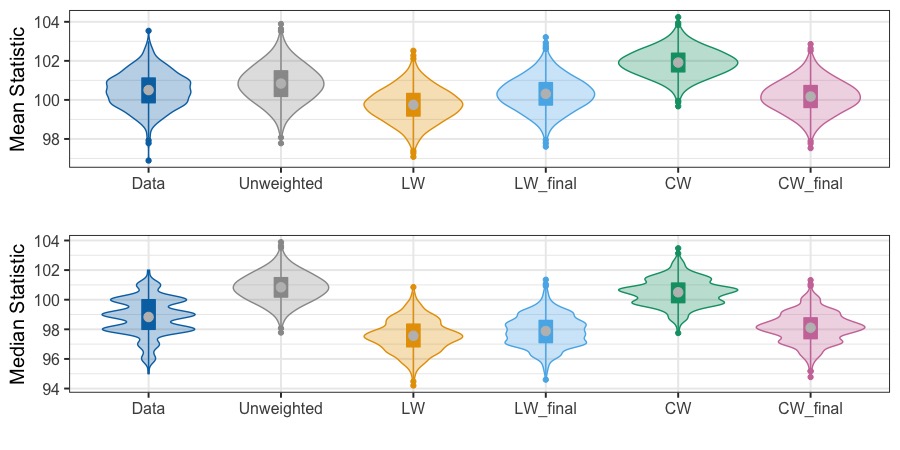}
      \label{fig:NB-utility1-reweight}}\\
    \subfloat[Posterior Density of $\mu$]{    \includegraphics[width=0.48\textwidth]{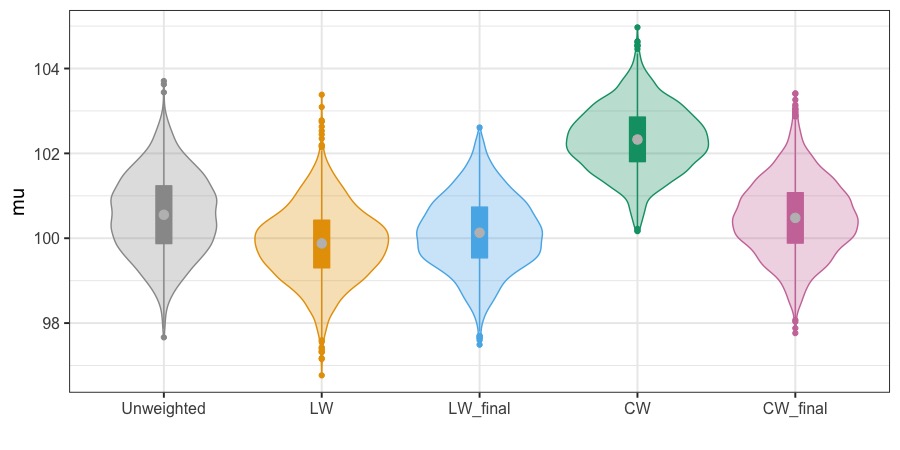}
    \label{fig:NB-mu-reweight}}
    \subfloat[Synthetic Data]{\includegraphics[width=0.48\textwidth]{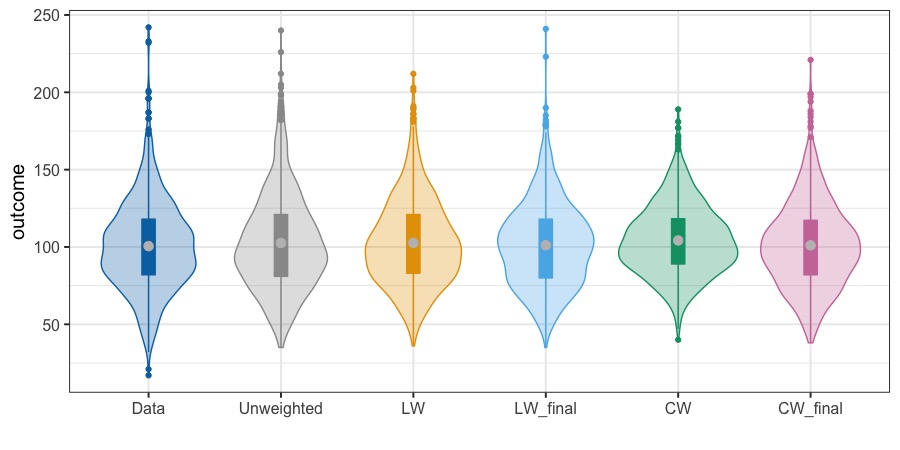}
      \label{fig:NB-y-reweight}}
    \caption{Before vs after re-weighting for the negative binomial mixture: utility. The re-weight strategy improves the utility of several quantities of interest for both LW and CW, with a more significant improvement for CW.}
    \label{fig:NB-reweight-utility}
\end{figure}

Turning to the utility performances, Figure \ref{fig:NB-utility4-reweight} and Figure \ref{fig:NB-utility1-reweight} show notable improvement in utility of CW after re-weighting for all estimates of the generated synthetic data. The deterioration in the preservation of the real (not synthetic) data distribution tails due to overly downweighting records in the tails \emph{before} re-weighting is greatly mitigated by the re-weighting strategy.  We observe that all of the extreme quantiles, the mean, and the median estimates are much more accurate. The improvement of utility of LW is less impressive because the relative improvement in the efficiency of the weighting scheme is relatively smaller. However, we can certainly see improvement in estimating the mean, median, and 90th quantile. For example, compared to a 95\% confidence interval of median in the data [96.0, 101.0], the CW\_final achieved [96.1, 100.1] improved from CW's [98.7, 102.4], and LW\_final achieved [96.0 100.0] improved from LW's [95.6, 99.7]. We include a table of comparisons of all estimands in the Supplementary Materials for brevity.

Moreover, utility of the mean parameter $\mu$ estimation in Figure \ref{fig:NB-mu-reweight} improves after re-weighting for LW and CW, with a bigger improvement for CW. When we turn to violin plot for the the data distribution, on the one hand, as compared to the synthesizer distributions, on the other hand, that is displayed in Figure \ref{fig:NB-y-reweight}, we see that the re-weighting strategy improves the preservation of the confidential data distribution tail in LW\_final and CW\_final, compared to LW and CW, respectively. This reduced down-weighting of the distribution tails is a feature of the re-weighting strategy for any vector-weighted scheme applied to highly skewed data.

\section{Application to The Survey of Doctorate Recipients}
\label{app}

The Survey of Doctorate Recipients (SDR) provides demographic, education, and career history information from individuals with a U.S. research doctoral degree in a science, engineering, or health (SEH) field. The SDR is sponsored by the National Center for Science and Engineering Statistics and by the National Institutes of Health. In this section, we demonstrate our re-weighting strategy on a sample of 1000 observations of the SDR focused on the highly skewed salary variable. The sample comes from the 2017 Survey of Doctorate Recipients public use file (\url{https://ncsesdata.nsf.gov/datadownload/}). The highly skewed salary variable has a mean of \$107,609, a median of a \$95,000, a range of [\$0, \$509,000], and a standard deviation of \$69,718. We use a negative binomial unweighted synthesizer for this highly skewed variable salary.

\subsection{Before re-weighting}
\label{app:before}
\emph{Before} re-weighting, the results of LW and CW on the real skewed data sample tell a similar story as on simulated skewed data. The results are included in the Supplementary Materials for brevity and we summarize the findings here: i) LW has the highest utility among the three, though its relatively heavy downweighting of records in the tails of the confidential data distribution under highly skewed data results in reduced utility compared to that of less skewed data; ii) CW's utility performance on the real skewed data is worse than that on the simulated skewed data--it has assigned low weights to many more records, resulting in low weighting efficiency and therefore low utility.

LW's weighting efficiency is close to optimal for most data records such that re-weighting  might not improve much. However, CW's weighting efficiency is expected to see huge improvement after re-weighting, which in turn, will improve its utility performance. 

\begin{figure}[t]
  \centering
    \includegraphics[width=1\textwidth]{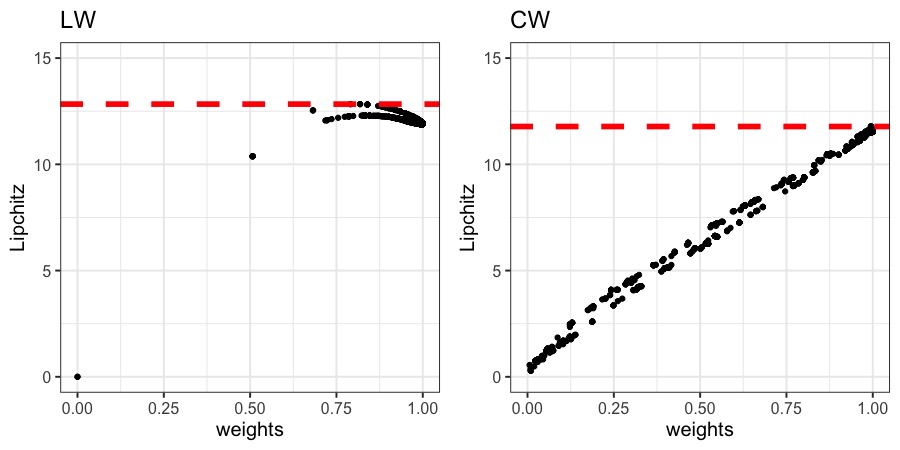}
    \caption{Lipschitz Bounds vs Weights, LW (left) and CW (right). The LW creates the majority of records expressing high Lipschitz bounds due to high weight values and therefore more efficient. The CW produces produces many records with low Lipscthiz bounds due to their over-downweighting.}
    \label{fig:app-L-alpha}
\end{figure}

\subsection{After re-weighting}
\label{app:after}

We apply the re-weighting strategy to LW and CW to maximize their utility performances. We set $k = 0.95$ to maintain an equivalent overall Lipschitz bound. The results are labeled as ``LW\_final" and ``CW\_final" respectively.

\begin{figure}[t]
    \centering
    \subfloat[LW]{    \includegraphics[width=0.48\textwidth]{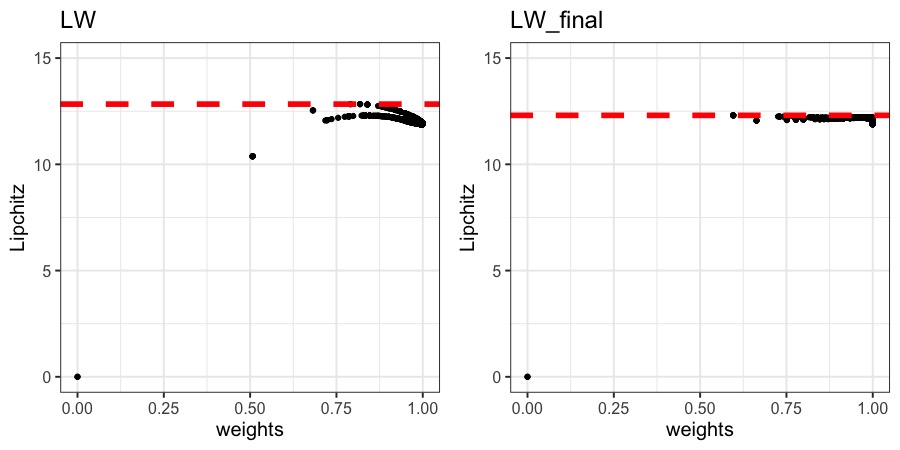}
    \label{fig:app-alpha-L-LW}}
    \subfloat[CW]{    \includegraphics[width=0.48\textwidth]{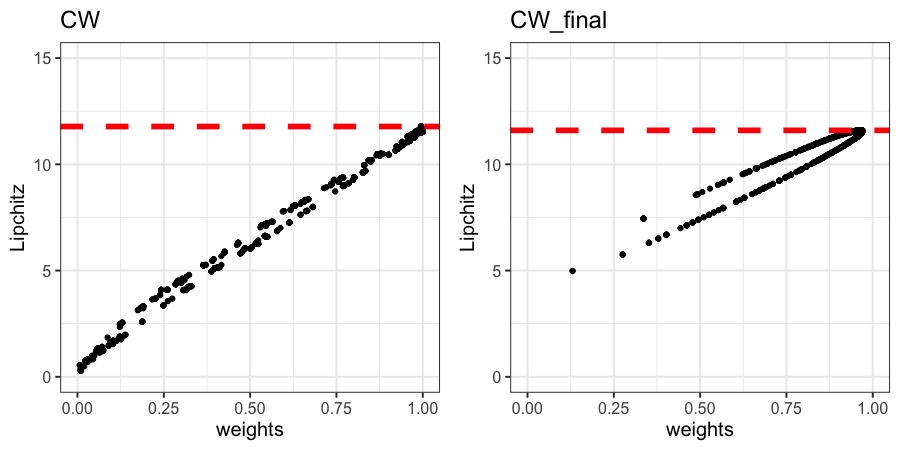}
    \label{fig:app-alpha-L-CW}}
    \caption{Lipschitz Bounds vs Weights, before and after re-weighting for LW (left) and CW (right). The re-weighting strategy produces a nearly horizontal Lipschitz-weight relationship. The impact is more significant for the LW.}
    \label{fig:app-alpha-L}
\end{figure}

Figure \ref{fig:app-alpha-L-LW} shows that the re-weighting strategy has pushed the Lipschitz-to-weight association to almost horizontal at the maximum Lipschitz bound,  $\Delta_{\bm{\alpha},\mathbf{x}}$, for LW, indicating maximum efficiency. Figure \ref{fig:app-alpha-L-CW} shows that the re-weighting strategy has also produced a  Lipschitz-to-weight association that is less vertical for CW. Therefore, we expect to see minor utility improvement of LW\_final, and a huge utility improvement of CW\_final.

\begin{figure}[t]
    \centering
    \subfloat[Lipschitz Bounds]{    \includegraphics[width=0.48\textwidth]{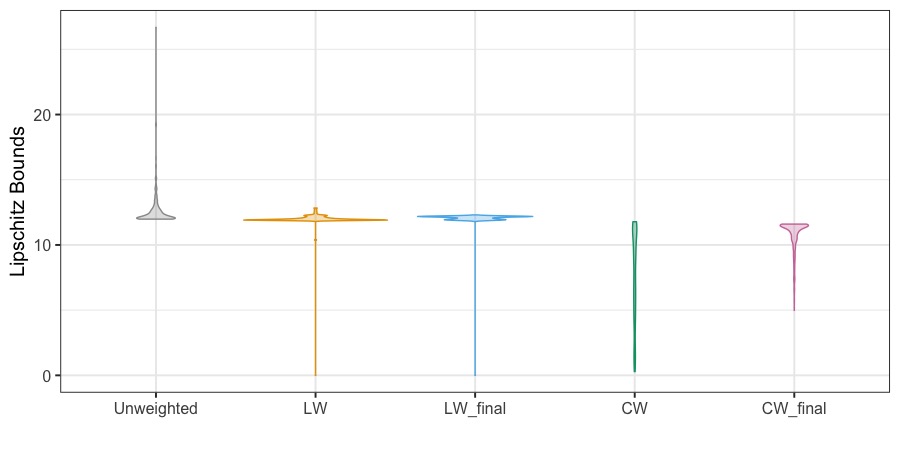}
    \label{fig:app-risks-2}}
    \subfloat[Record-level Weights]{    \includegraphics[width=0.48\textwidth]{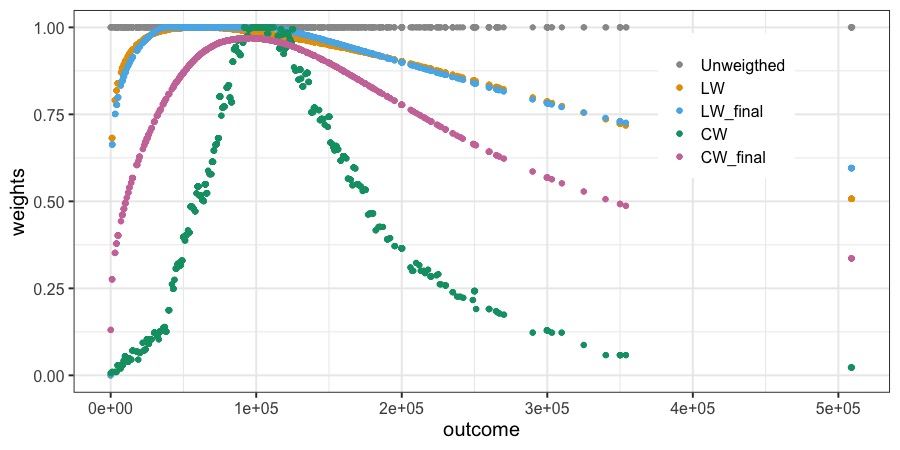}
    \label{fig:app-weights-dots-2}}\\
    \subfloat[LW Weights]{    \includegraphics[width=0.48\textwidth]{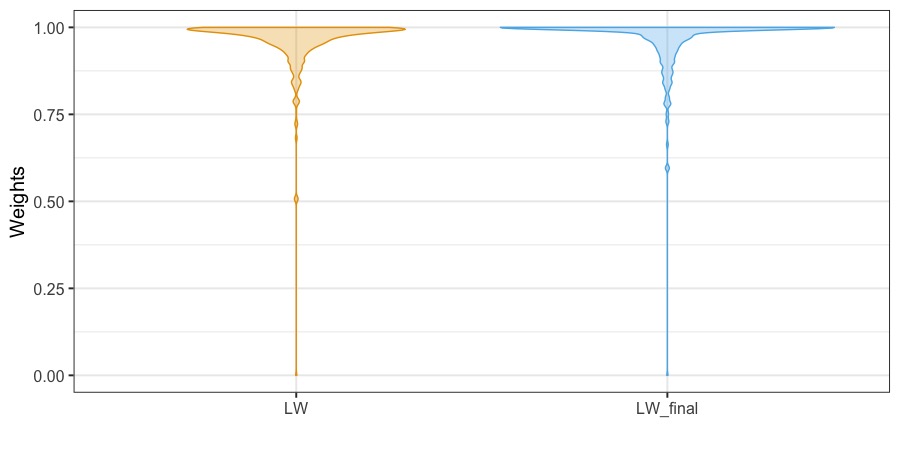}
    \label{fig:app-weights-LW}}
    \subfloat[CW Weights]{    \includegraphics[width=0.48\textwidth]{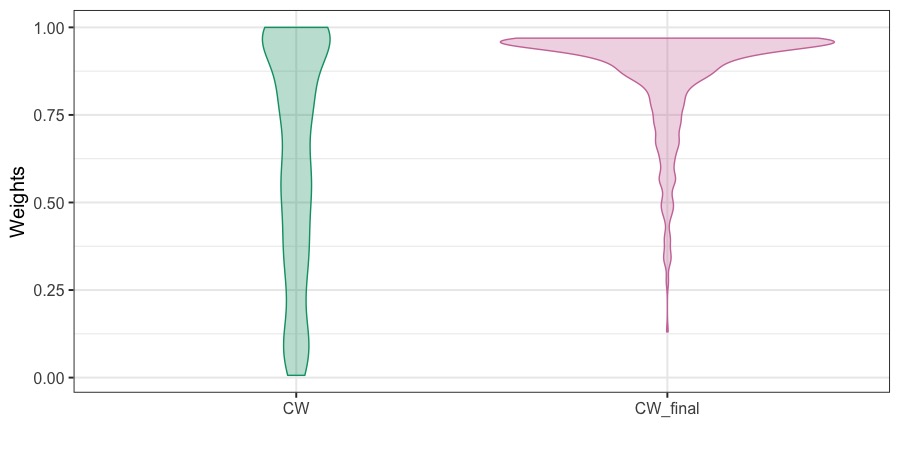}
    \label{fig:app-weights-CW}}
    \caption{Before vs after re-weighting for salary: Lipschitz Bounds (panel a) and Weights (panels b to d). After re-weighting, the by-record Lipschitz bounds are closer to the maximum, whereas the record-level weights are closer to one, indicating the effectiveness of the re-weighting strategy to increase the weighting efficiency of both LW and CW.}
    \label{fig:app-reweight-weights}
\end{figure}

Examining the distributions for the weights under both of LW and CW in Figure \ref{fig:app-reweight-weights} shows how much weighting efficiency LW\_final and CW\_final have gained after the re-weighting strategy. Focusing on the walk between CW to CW\_final in Figure \ref{fig:app-weights-CW} the distribution of weights is highly diffuse with large mass assigned to low values before re-weighting.  After re-weighting, by contrast, many more records receive higher weights. Even though the by-record weights ($\bm{\alpha}$) have increased after re-weighting, Figure \ref{fig:app-risks-2} shows that the re-weighting strategy has maintained an equivalent maximum Lipschitz bound $\Delta_{\bm{\alpha},\mathbf{x}}$.

\begin{figure}[H]
    \centering
    \subfloat[15th and 90th Quantiles]{    \includegraphics[width=0.48\textwidth]{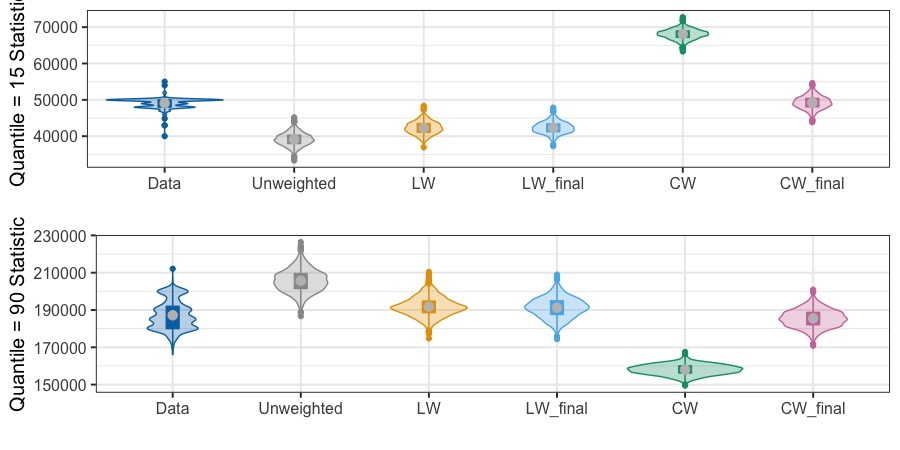}
      \label{fig:app-utility4-reweight}}
    \subfloat[Mean and Median]{    \includegraphics[width=0.48\textwidth ]{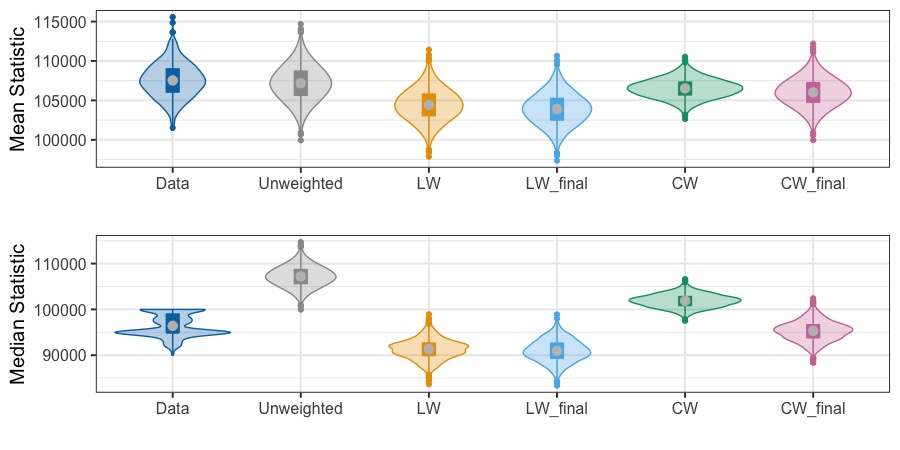}
      \label{fig:app-utility1-reweight}}\\
    \subfloat[Posterior Density of $\mu$]{    \includegraphics[width=0.48\textwidth]{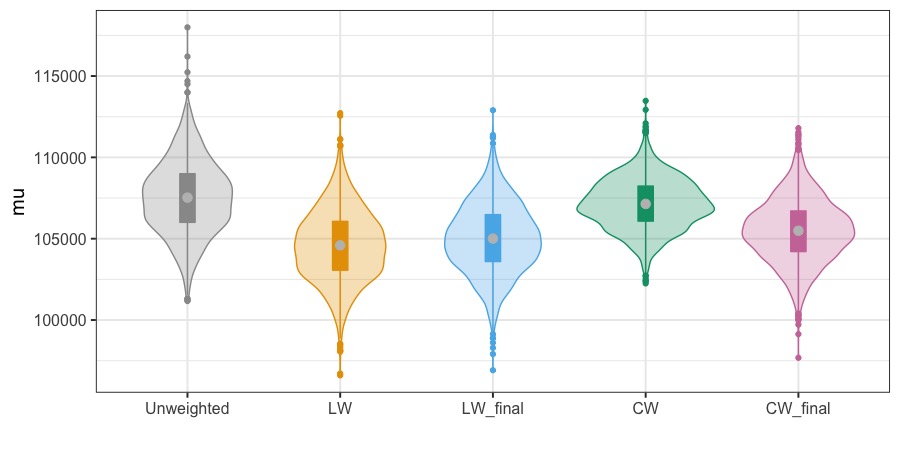}
    \label{fig:app-mu-reweight}}
    \subfloat[Synthetic Data]{\includegraphics[width=0.48\textwidth]{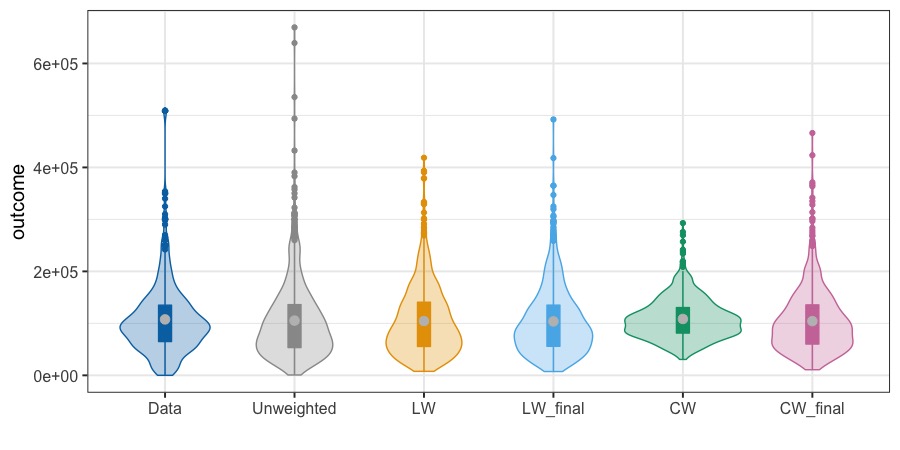}
      \label{fig:app-y-reweight}}
    \caption{Before vs after re-weighting for salary: utility. The re-weight strategy improves the utility of several quantities of interest for both LW and CW, with a more significant improvement for CW.}
    \label{fig:app-reweight-utility}
\end{figure}

Finally, the utility results in Figure \ref{fig:app-reweight-utility} demonstrate the utility maximizing feature of our proposed re-weighting strategy on the confidential data sample. Whether it is the preservation of statistics of the confidential data distribution, shown in Figure \ref{fig:app-utility4-reweight} and Figure \ref{fig:app-utility1-reweight}, or the relative accuracy of parameter estimates in Figure \ref{fig:app-mu-reweight}, or similarity of the synthetic data density to that of the confidential data in Figure \ref{fig:app-y-reweight}, CW\_final has produced much higher utility than CW across the board, a result that we expect to see given its improved weighting efficiency previously discussed. We make particular mention that the comparisons of the confidential and synthetic data distributions in Figure~\ref{fig:app-y-reweight} show that re-weighting reduces or mitigates the shrinking of the tail of the confidential data in the resulting synthetic data. Overall, there is a minor utility improvement from LW to LW\_final, another result we expect to see given its minor improvement of weighting efficiency. Nevertheless, our proposed re-weighting strategy maximizes the utility of any vector-weighted scheme, while maintaining an equivalent maximum Lipschitz bound $\Delta_{\bm{\alpha},\mathbf{x}}$.

\section{Concluding Remarks}
\label{conclusion}

In this article, we introduce a new re-weighting strategy that improves utility of \emph{any} vector-weighted scheme in the difficult case of a highly-skewed data distribution, while maintaining an equivalent privacy budget. Applied to both the LW and CW vector-weighted synthesizers, this strategy improves their weighting efficiency by increasing by-record weights to compress the distribution of by-record Lipschitz bounds. Improved weighting efficiency substantially mitigates the tendency for vector-weighted schemes to overly downweight the tails, especially for CW. 



\bibliography{DPbib}
\bibliographystyle{natbib}

\section*{Supp1: R scripts of Algorithm 1 in Section 2.1}

\subsection*{\bf Computing weights $\bm{\alpha}$:}

\noindent The \texttt{stan\_estimate\_unweighted} below is the Stan output of the unweighted synthesizer.

\begin{verbatim}
## step 1
log_lik  <- stan_estimate_unweighted$log_lik ## S x N matrix
N  <- ncol(log_lik)
S <- nrow(log_lik)
log_ratio <- matrix(0,S,N)
log_ratio_theta <- matrix(0,S,1)
pos <- rep(TRUE,N)

## step 2
for( s in 1:S ){
    log_like_xs  <- sum(log_lik[s,]) ## full data
    for(i in 1:N){
      pos_i <- pos
      pos_i[i] <- FALSE
      log_like_xsi <- sum(log_lik[s,pos_i])
      log_ratio[s,i] <- abs(log_like_xs - log_like_xsi)
    }}
log_ratio_theta <- rowMaxs(log_ratio, value = TRUE)
L  <- quantile(log_ratio_theta,thresh)

## step 3
log_ratio_data  <- colMaxs(logthresh_ratio,value=TRUE)
f_linres <- function(x){(x-min(x))/(max(x)-min(x))}
risks <- f_linres( log_ratio_data )

## step 4
weights <- c * (1 - risks) + g
weights[weights <= 0] <- 0
weights[weights >= 1] <- 1
\end{verbatim}

\subsection*{\bf Compute Lipschitz bound, $\Delta_{\bm{\alpha},\mathbf{x}}$:}

\noindent The \texttt{stan\_estimate\_weighted} below is the Stan output of the weighted synthesizer. Step 5 is done by Stan estimation.

\begin{verbatim}
## step 6
log_lik  <- stan_estimate_weighted$log_lik ## S x N matrix
N  <- ncol(log_lik)
S <- nrow(log_lik)
log_ratio <- matrix(0,S,N)
log_ratio_theta <- matrix(0,S,1)
pos <- rep(TRUE,N)
for( s in 1:S ){
    log_like_xs  <- sum(log_lik[s,]) ## full data
    for(i in 1:N){
      pos_i <- pos
      pos_i[i] <- FALSE
      log_like_xsi <- sum(log_lik[s,pos_i])
      log_ratio[s,i] <- abs(log_like_xs - log_like_xsi)
    }}

## step 7
log_ratio_theta <- rowMaxs(log_ratio, value = TRUE)
L  <- quantile(log_ratio_theta,thresh)
\end{verbatim}

\section*{Supp 2: Stan script for a weighted negative binomial synthesizer}

\begin{verbatim}
functions{

real wt_NB_lpmf(int[] y, vector mu, real phi, vector weights, int n){
    real check_term;
    check_term  = 0.0;
    for( i in 1:n )
    {
	check_term    += weights[i] * neg_binomial_2_log_lpmf(y[i] | mu[i], phi);
    }
    return check_term;
  }

real wt_NBi_lpmf(int y_i, real mu_i, real phi, real weights_i){
    real check_term;
	check_term    = weights_i * neg_binomial_2_log_lpmf(y_i | mu_i, phi);
    return check_term;
  }
} 

data {
    int<lower=1> n; 
	  int<lower=1> K; 
    int<lower=0> y[n]; 
    vector<lower=0>[n] weights; 
    matrix[n, K] X; 
}

transformed data{
  vector<lower=0>[K] zeros_beta;
  zeros_beta  = rep_vector(0,K);
} 

parameters{
  vector[K] beta; 
  vector<lower=0>[K] sigma_beta; 
  cholesky_factor_corr[K] L_beta; 
  real reciprocal_phi; 
}

transformed parameters{
  vector[n] mu;
  real phi;
  mu   = X * beta;
  phi = 1. / reciprocal_phi;
} 

model{
  reciprocal_phi ~ cauchy(0., 5);
  L_beta          ~ lkj_corr_cholesky(6);
  sigma_beta      ~ student_t(3,0,1);
  beta            ~ multi_normal_cholesky( zeros_beta, diag_pre_multiply(sigma_beta,L_beta) ); 
  target          += wt_NB_lpmf(y | mu, phi, weights, n);
} 

generated quantities{
  vector[n] log_lik;
  for (i in 1:n) {
	log_lik[i] = wt_NBi_lpmf(y[i]| mu[i], phi, weights[i]);
	}
}
\end{verbatim}

\section*{Supp 3: Utility comparison before and after re-weighting in Section 3.2} 
Table \ref{tab:NB-utility} presents utility comparison of LW and CW before and after re-weighting.

\begin{table}[H]
\begin{center}
\begin{tabular}{l | c | c c | c c}
\hline
 & Data & LW & LW\_final & CW & CW\_final \\ \hline
 15th quantile & [70.0, 75.0] & [70.0, 74.3] & [70.6, 74.8] & [77.2, 80.8] & [71.6, 75.8]  \\
 90th quantile & [132.0, 140.0] & [132.1, 139.0] & [133.1, 140.3] & [128.5, 134.0] & 131.0, 137.7]\\
 mean & [98.8, 102.3] & [98.0, 101.4] & [98.7, 102.0] & 100.6, 103.3] & [98.6, 101.8]\\
 median & [96.0, 101.0] & [95.6, 99.7] & [96.0, 100.0] & [98.7, 102.4] & [96.1, 100.1] \\ \hline
\end{tabular}
\caption{Utility comparison before and after re-weighting for the negative binomial mixture: 95\% confidence interval}
\label{tab:NB-utility}
\end{center}
\end{table}

\section*{Supp 4: Plots of LW, CW, and SW before re-weighting in Section 4.1}

\begin{figure}[H]
  \centering
     \subfloat[Weights]{ \includegraphics[width=0.48\textwidth]{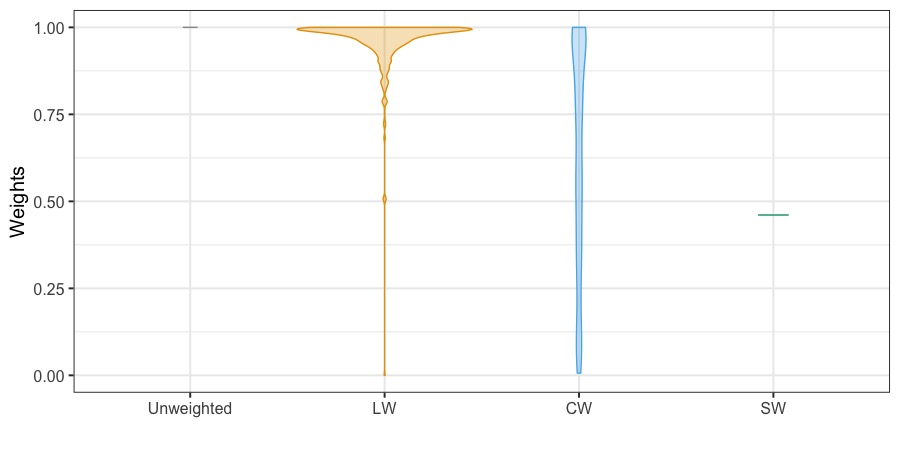}
    \label{fig:app-weights}}
  \subfloat[Lipschitz Bounds]{\includegraphics[width=0.48\textwidth]{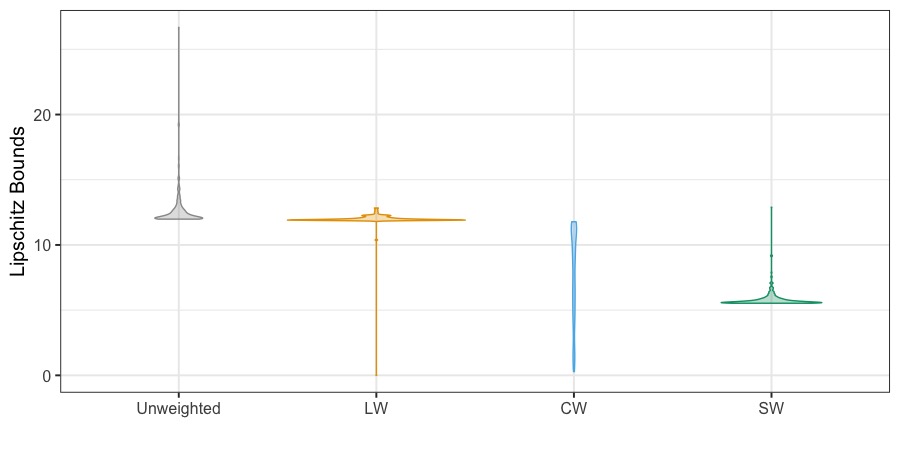}
    \label{fig:app-risks}}\\
  \subfloat[15th and 90th Quantiles]{\includegraphics[width=0.48\textwidth]{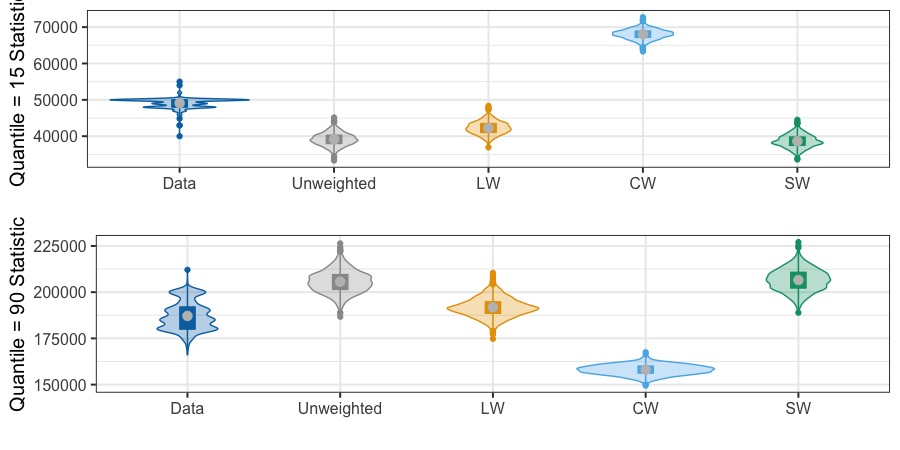}
      \label{fig:app-utility4}}
  \subfloat[Mean and Median]{\includegraphics[width=0.48\textwidth]{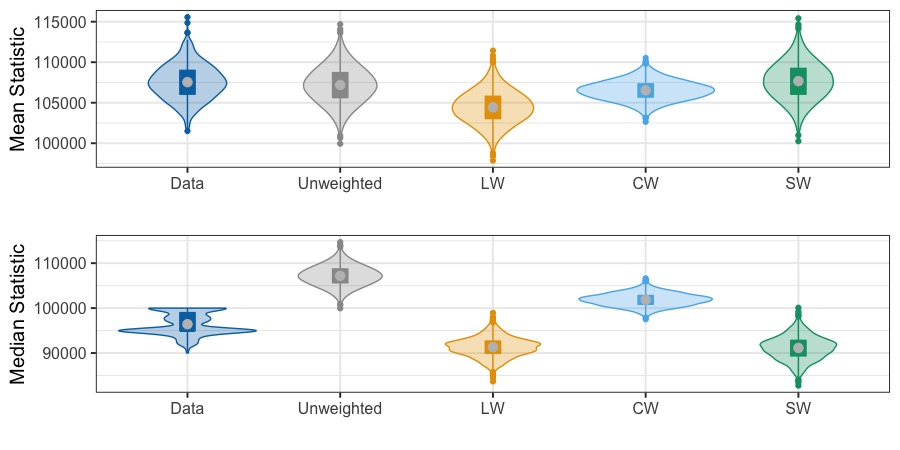}
      \label{fig:app-utility1}}
    \\
  \subfloat[Posterior Density of $\mu$]{
    \includegraphics[width=0.48\textwidth]{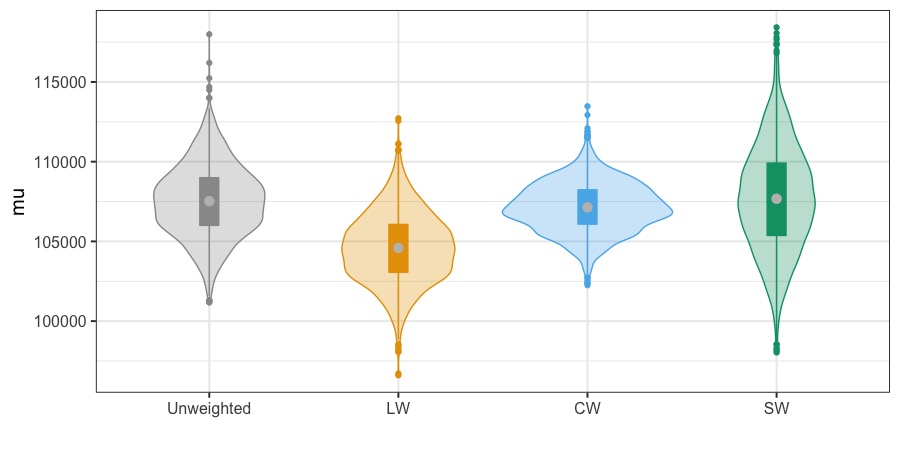}
    \label{fig:app-mu}}
     \subfloat[Synthetic Data]{
        \includegraphics[width=0.48\textwidth]{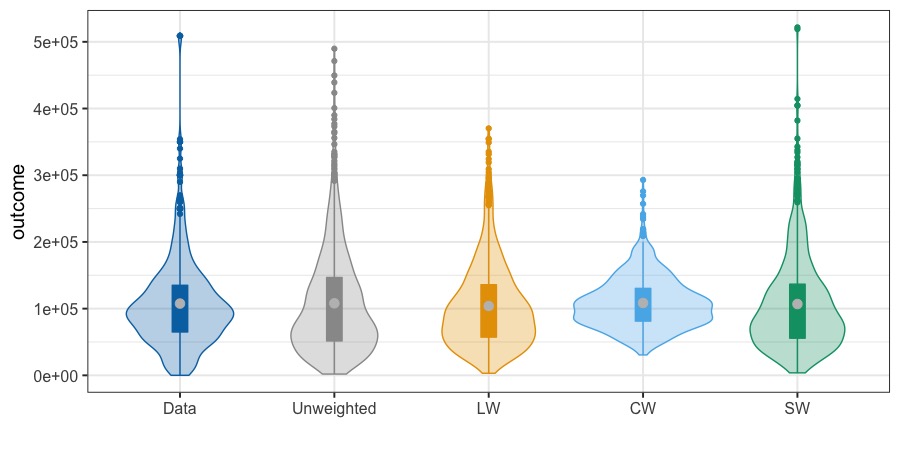}\label{fig:app-y}
     }
    \caption{Violin Plots for Salary}
    \label{fig:app-violin}
\end{figure}

Figure \ref{fig:app-violin} provides results of LW, CW, and SW \emph{before} re-weighting.

\section*{Supp 5: Moving from Local-to-Global Privacy Guarantee}
\label{mcsim}

We proceed to implement a Monte Carlo simulation study under each of our less skewed Poisson and more skewed mixture of negative binomials data generating models to walk from the local privacy guarantee for a specific database, $\mathbf{x}$, to a global (asymptotic DP) guarantee over the space of databases, $\forall\mathbf{x}\in\mathcal{X}^n$. We generate $R = 100$ local databases under each generating model, estimate the unweighted and LW weighted synthesizers on each database, $r \in \left(1,\ldots,(R=100)\right)$, and compute a local Lipschitz bound, $\Delta_{\bm{\alpha},\mathbf{x}_{r}}$, on each $\mathbf{x}_{r}$ under each synthesizer.  We plot the distributions of the $(\Delta_{\bm{\alpha},\mathbf{x}_{r}})_{r=1}^{R = 100}$ for each synthesizer and conclude that we have achieved a global asymptotic DP result if this distribution contracts around a global $\Delta_{\bm{\alpha}}$.  We summarize our Monte Carlo simulation procedure, below:

\begin{enumerate}
\item  For $r = 1,\ldots, (R=200)$:
\begin{itemize}
	\item Generate $\mathbf{x}_r \sim \text{Pois}(\mu)$ or $\mathbf{x}_r \sim \pi_{1}\text{NB}(\mu_{1}=100,\phi_{1} = 5) + \pi_{2}\text{NB}(\mu_{2}=100,\phi_{2} = 5)$, each of size $n = 1000$.
	\item Compute the \emph{local} Lipschitz bound, $\Delta_{\bm{\alpha},\mathbf{x}_{r}}$, for the unweighted and $\bm{\alpha}-$re-weighted synthesizers.
    \item  Construct the distribution of $\Delta_{\bm{\alpha},\mathbf{x}_{r}}$ and note the maximum of the distribution and difference between the maximum and minimum values of the distribution of the local Lipschitz bounds.
\end{itemize}
\item Assess contraction of the $\max_{r}\Delta_{\bm{\alpha},\mathbf{x}_{r}}$ to a single (global) value and whether the minimum and maximum values collapse together.
\end{enumerate}

\begin{figure}[H]
    \centering
    \includegraphics[width=1\textwidth]{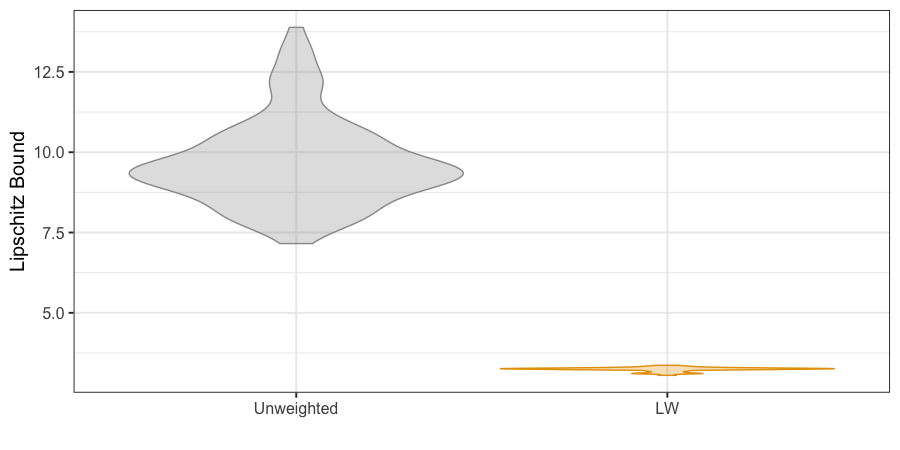}
    \caption{Violin plot of Lipschitz bounds over the Monte Carlo iterations under the Poisson generating model.}
    \label{fig:Poisson-MC-bounds}
\end{figure}

Figure~\ref{fig:Poisson-MC-bounds} presents a violin plot of the local $(\Delta_{\bm{\alpha},\mathbf{x}_{r}})_{r=1}^{R}$ for the $R = 100$ Monte Carlo iterations of the less skewed data generating model for the Unweighted (left) and LW-weighed (right) synthesizers.  We readily observe that the LW synthesizer contracts onto the global value, $\epsilon = 2\Delta_{\bm{\alpha}} = 7$.  That this contraction is consistent with a relaxed, pDP guarantee comes from the small distribution mass above $\Delta_{\bm{\alpha}} = 3.5$, though we see the probability that the Lipschitz bound any local database exceeds $3.5$ is nearly $0$ at $n = 1000$.

\begin{figure}[H]
    \centering
    \subfloat[Mean and Median]{   \includegraphics[width=0.48\textwidth]{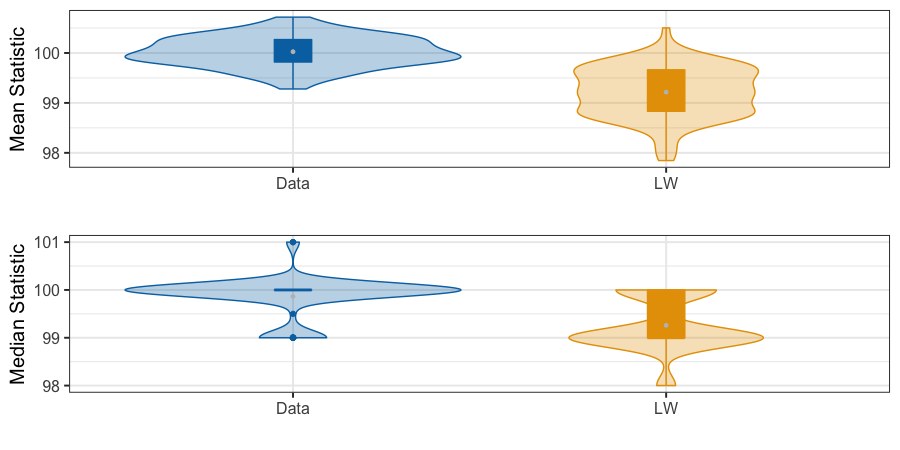}
      \label{fig:poisson-MC-mean-median}}
    \subfloat[15th and 90th Quantiles]{    \includegraphics[width=0.48\textwidth]{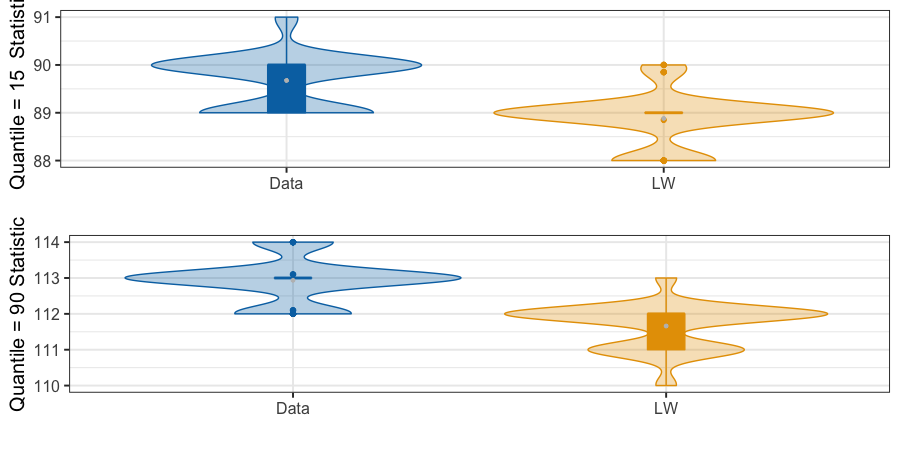}
    \label{fig:poisson-MC-q15-q90}}
    \caption{Violin Plots for the estimands of the confidential and synthetic data distributions over the Monte Carlo iterations for the Poisson generating model.}
    \label{fig:Poisson-MC-utility}
\end{figure}

Figure~\ref{fig:Poisson-MC-utility} presents the associated distributions of a set of estimands over the Monte Carlo iterations for each synthesizer as compared to the confidential data.  There is an expected loss of utility, though inference is reasonably well-preserved.

The following set of figures repeat the earlier set, but here under the highly skewed data generation process from a negative binomial mixture.  The conclusions are the same that we observe substantial contraction around the global Lipschitz (where here $\Delta_{\bm{\alpha}} = 6$).

\begin{figure}[H]
    \centering
    \includegraphics[width=1\textwidth]{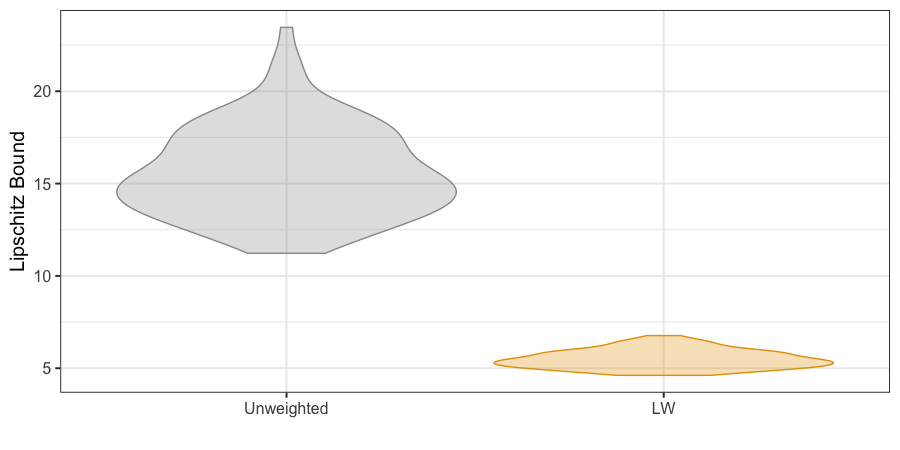}
    \caption{Violin plot of Lipschitz bounds over the Monte Carlo iterations under the mixtures of negative binomials generating model.}
    \label{fig:NB-MC-bounds}
\end{figure}

\begin{figure}[H]
    \centering
    \subfloat[Mean and Median]{  
    \includegraphics[width=0.48\textwidth]{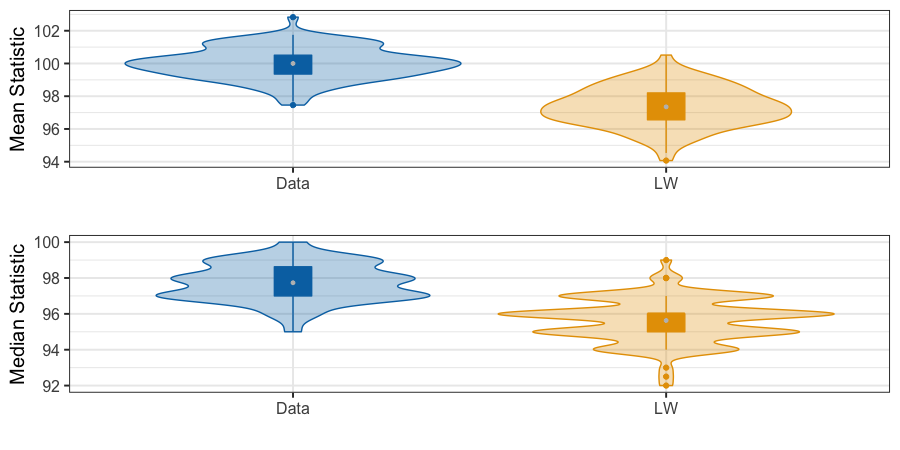}
      \label{fig:NB-MC-Lbounds}}
    \subfloat[15th and 90th Quantiles]{    \includegraphics[width=0.48\textwidth]{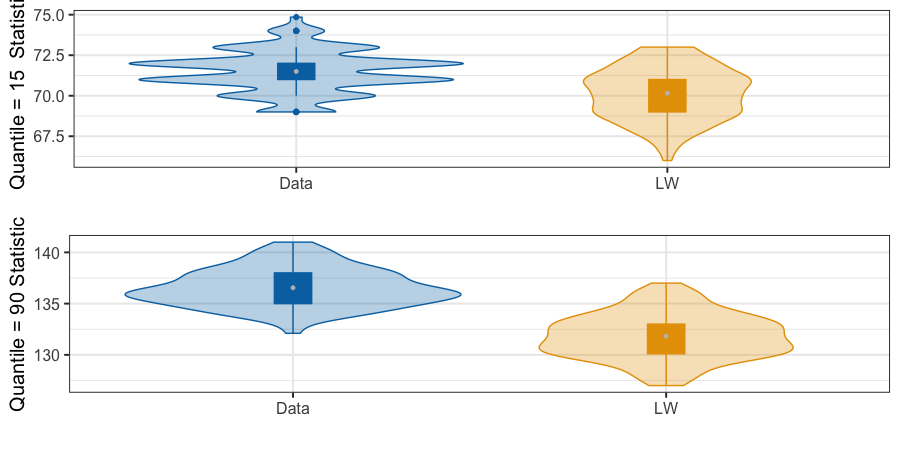}
    \label{fig:NB-MC-q15-q90}}
    \caption{Violin Plots for the estimands of the confidential and synthetic data distributions over the Monte Carlo iterations for the mixtures of negative binomials generating model.}
    \label{fig:NB-MC-utility}
\end{figure}

We have illustrated the theory of \citet{SavitskyWilliamsHu2020ppm} that guarantees an asymptotic DP result by demonstrating a contraction of a collection of Lipschitz bounds for local databases onto a global Lipschitz bound under both of our low and highly skewed data generating scenarios, with our LW synthesizer and re-weighting strategy.   The key conclusion is that for $n$ sufficiently large that a local Lipschitz bound estimated on a specific database, $\mathbf{x}$, becomes arbitrarily close to the global Lipschitz bound.

\end{document}